\documentclass[twoside,twocolumn,9pt]{article}
\usepackage{extsizes}
\usepackage[super,sort&compress,comma]{natbib} 
\usepackage[version=3]{mhchem}
\usepackage[left=1.5cm, right=1.5cm, top=1.785cm, bottom=2.0cm]{geometry}
\usepackage{balance}
\usepackage{mathptmx}
\usepackage{sectsty}
\usepackage{graphicx} 
\usepackage{lastpage}
\usepackage[format=plain,justification=justified,singlelinecheck=false,font={stretch=1.125,small,sf},labelfont=bf,labelsep=space]{caption}
\usepackage{float}
\usepackage{fancyhdr}
\usepackage{fnpos}
\usepackage[english]{babel}
\addto{\captionsenglish}{%
  
}
\usepackage{array}
\usepackage{droidsans}
\usepackage{charter}
\usepackage[T1]{fontenc}
\usepackage[usenames,dvipsnames]{xcolor}
\usepackage{setspace}
\usepackage[compact]{titlesec}
\usepackage{hyperref}

\usepackage{soul}

\usepackage{braket}

\definecolor{cream}{RGB}{222,217,201}

\newcommand{\ems}[1]{\textcolor{black}{#1}}

\newcommand{\mlm}[1]{\textcolor{black}{#1}}

\begin{document}

\pagestyle{fancy}
\thispagestyle{plain}
\fancypagestyle{plain}{
\renewcommand{\headrulewidth}{0pt}
}

\makeFNbottom
\makeatletter
\renewcommand\LARGE{\@setfontsize\LARGE{15pt}{17}}
\renewcommand\Large{\@setfontsize\Large{12pt}{14}}
\renewcommand\large{\@setfontsize\large{10pt}{12}}
\renewcommand\footnotesize{\@setfontsize\footnotesize{7pt}{10}}
\makeatother

\renewcommand{\thefootnote}{\fnsymbol{footnote}}
\renewcommand\footnoterule{\vspace*{1pt}%
\color{cream}\hrule width 3.5in height 0.4pt \color{black}\vspace*{5pt}} 
\setcounter{secnumdepth}{5}

\makeatletter 
\renewcommand\@biblabel[1]{#1}            
\renewcommand\@makefntext[1]%
{\noindent\makebox[0pt][r]{\@thefnmark\,}#1}
\makeatother 
\renewcommand{\figurename}{\small{Fig.}~}
\sectionfont{\sffamily\Large}
\subsectionfont{\normalsize}
\subsubsectionfont{\bf}
\setstretch{1.125} 
\setlength{\skip\footins}{0.8cm}
\setlength{\footnotesep}{0.25cm}
\setlength{\jot}{10pt}
\titlespacing*{\section}{0pt}{4pt}{4pt}
\titlespacing*{\subsection}{0pt}{15pt}{1pt}

\fancyfoot{}
\fancyhead{}
\renewcommand{\headrulewidth}{0pt} 
\renewcommand{\footrulewidth}{0pt}
\setlength{\arrayrulewidth}{1pt}
\setlength{\columnsep}{6.5mm}
\setlength\bibsep{1pt}

\makeatletter 
\newlength{\figrulesep} 
\setlength{\figrulesep}{0.5\textfloatsep} 

\newcommand{\topfigrule}{\vspace*{-1pt}%
\noindent{\color{cream}\rule[-\figrulesep]{\columnwidth}{1.5pt}} }

\newcommand{\botfigrule}{\vspace*{-2pt}%
\noindent{\color{cream}\rule[\figrulesep]{\columnwidth}{1.5pt}} }

\newcommand{\dblfigrule}{\vspace*{-1pt}%
\noindent{\color{cream}\rule[-\figrulesep]{\textwidth}{1.5pt}} }

\makeatother

\twocolumn[
  \begin{@twocolumnfalse}
\sffamily

\noindent\LARGE{\textbf{Avalanche dynamics in sheared athermal particle packings occurs via localized bursts predicted by unstable linear response}}
  
\vspace{0.3cm}
  
\noindent\large{Ethan Stanifer\textit{$^{*}$} and M. Lisa Manning\textit{$^{\ddag}$}}\\

\noindent\normalsize{Under applied shear strain, granular and amorphous materials deform via particle rearrangements, which can be small and localized or organized into system-spanning avalanches. While the statistical properties of avalanches under quasi-static shear are well-studied, the dynamics during avalanches is not. In numerical simulations of sheared soft spheres, we find that avalanches can be decomposed into bursts of localized deformations, which we identify using an extension of persistent homology methods. We also study the linear response of unstable systems during an avalanche, demonstrating that eigenvalue dynamics are highly complex during such events, and that the most unstable eigenvector is a poor predictor of avalanche dynamics. Instead, we modify existing tools that identify localized excitations in stable systems, and apply them to these unstable systems with non-positive definite Hessians, quantifying the evolution of such excitations during avalanches. We find that bursts of localized deformations in the avalanche almost always occur at localized excitations identified using the linear spectrum. These new tools will provide an improved framework for validating and extending mesoscale elastoplastic models that are commonly used to explain avalanche statistics in glasses and granular matter.}

 \end{@twocolumnfalse} \vspace{0.6cm}

  ]


\renewcommand*\rmdefault{bch}\normalfont\upshape
\rmfamily
\section*{}
\vspace{-1cm}


\footnotetext{\textit{$^{*}$~Department of Physics, University of Michigan, Ann Arbor, Michigan 48109, USA}}
\footnotetext{\textit{$^{\ddag}$~Department of Physics and BioInspired Institute, Syracuse University, Syracuse, New York 13244, USA}}
\footnotetext{\textit{$^{\ddag}$~To whom correspondence should be addressed: mmanning@syr.edu}}




\section{Introduction}

Can we predict how amorphous materials such as bulk metallic glasses~\cite{wang2004bulk,SCHUH2007}, dense colloidal suspensions~\cite{schall2007structural}, and foams~\cite{Lauridsen2002,Bonn2017} fail under stress? While most foams and crystalline metals are ductile and fail homogeneously, bulk metallic glasses fail catastrophically via shear bands or system-spanning avalanches~\cite{wang2004bulk,Zhang2006}, despite their otherwise desirable material properties. What microscopic properties generate this macroscopic difference in yielding behavior? Such a fundamental description would allow rational material design to control failure mechanisms such as shear bands and avalanches~\cite{SCHUH2007}.
	


Unfortunately, it remains unclear what micro- or meso-scopic features govern this brittle-to-ductile transition. Previous work on athermal avalanches have largely focused on systems under athermal quasistatic shear, where configurations are analyzed before and after the system spanning rearrangements~\cite{Lematre2009,Salerno2012}. A few works have also focused on packings sheared under finite strain rate~\cite{Salerno2012, nicolas2014, lagogianni2018}. These studies evaluate the size, statistics, and/or shape distribution of these avalanches as a function of material properties, often with a focus on the ductility of the initial configuration~\cite{Salerno2012,Maloney2006}.


Phenomenological work has focused on understanding the transition from ductile to brittle failure in terms of macroscopic system parameters such as composition, temperature, or preparation~\cite{Nicolas2017,Ozawa2018,Popovi2018}. 
Recently some authors have used mesoscopic elastoplastic models to investigate the origin of the transition from a brittle-to-ductile behavior~\cite{Ozawa2018,Popovi2018}. In these models, it is assumed the system is comprised of independent, mesoscopic yielding regions and that the stress to yield $x$ in each region is taken from a specified distribution $P_0(x)$ that captures different preparation protocols. In poorly annealed systems, the mean of $P_0(x)$  is expected to be small, while in well-annealed systems it is large. This hypothesis is strongly supported by work from Patinet et al.~\cite{Patinet2016} who explicitly measure local yield stresses, with some assumptions and caveats, in simulated granular systems. Other models such as Shear Transformation Zone (STZ) and Soft Glassy Relaxation (SGR) also describe localized regions that deform and fail within the glassy systems~\cite{falk1998dynamics, Manning2007,sollich2006soft}. 

At a microscopic level these models assume that plasticity is controlled by "shear transformations", the discrete localized events where a small number of particles rearrange locally, which release the accumulated stress~\cite{Barlow,sollich2006soft}. \mlm{This is corroborated by several manuscripts showing that the full displacement field of an avalanche is well-fit by a set of Eshelby transformations, each centered in a localized region of high deformation~\cite{dasgupta2013yield, albaret2016mapping}.}
Similar to elastoplastic models~\cite{Popovi2018, ozawa2018random}, this implies system-spanning rearrangements occur as sequential bursts of localized motion \mlm{although the temporal dynamics of these localized bursts have not been well-studied.  Such dynamics are important, as the models differ in their predictions for} how defects are coupled dynamically during an avalanche. Elastoplastic models couple defects by explicitly quadrapolar elastic stress fields while the STZ/SGR models couple defects via local structural changes and noise. In order to test these predictions for coupling between soft regions during avalanches, we first need a robust method for extracting \mlm{the temporal dynamics of} soft regions from unstable amorphous packings.

Unlike dislocations in crystalline solids, defects in amorphous solids are not easily identified by the local geometry.
One way to find soft, defect-like regions is via direct measurements of local yield stress by Patient and collaborators~\cite{Patinet2016} discussed above. Additionally, quite a few new techniques have been developed to identify structural indicators that predict plasticity. Some focus on the linear or nonlinear response near an instability ~\cite{manning2011vibrational,zylberg2017thermal,Wijtmans2017,Ding2014,gartner2016nonlinear,richard2021simple}, while others use machine learning techniques~\cite{schoenholz2016structural, bapst2020unveiling} or identification of high energy motifs~\cite{tong2018revealing}. Recently, many of these have been compared on the same set of data across the brittle-to-ductile transition, and one conclusion of that work is that structural defect indicators based on linear response are surprisingly predictive~\cite{Richard2020}.

Linear response indicators are computed from the hessian: $H = \partial^2 U/ \partial u_i \partial u_j$, where $U$ is the potential energy and $u_i$ is the displacements of particle $i$~\cite{ellenbroek2007response}.
One such structural indicator is simply a weighted superposition of the lowest energy eigenmodes of the dynamical matrix, or vibrational modes.  The resulting field is termed "vibrality"~\cite{tong2014}, or "soft spots" if the field is clustered~\cite{manning2011vibrational}.  Of course, the lowest energy vibrational mode just before the instability is precisely the initial motion during the avalanche~\cite{manning2011vibrational,rottler2014predicting}, and the nonaffine displacement field is dominated by these low-energy modes close to the instability, as well~\cite{Richard2020,manning2011vibrational,Wijtmans2017}. 

However, structural metrics based on linear response may fail to predict large avalanches because they are computed once before the avalanche, and can not be computed during the avalanche.  That is because these methods assume a positive-definite dynamical matrix, but during most of the avalanche dynamics, the Hessian matrix has at least one negative eigenvalue. Therefore, it is obvious to ask whether some of the methods for identifying soft spots in positive-definite Hessians can be extended to Hessians describing unstable systems.

In this manuscript, we develop such extensions and calculate soft spots in order to investigate how they evolve over the course of an avalanche. To understand whether we can really describe avalanches as bursts of localized motions, we also develop a new method for isolating non-affine movements in the $D^2_{min}$ field~\cite{falk1998dynamics} using an extension of persistent homology. This allows us to robustly separate an avalanche into a set of localized rearrangements. Finally, we compare these rearrangements to evolving soft spots to understand how soft spots are coupled to generate the observed dynamics. 

These methods may be useful not only for quasistatically sheared athermal systems, but potentially many other unstable systems such as active matter systems, which may be amenable to similar techniques, or thermal systems which are typically not in mechanical equilibrium.

\section{Methods}
\subsection{Simulating dynamics of the sheared granular packing}
\label{secdyn}
We study bidisperse granular packings. Particles interact with a Hertzian contact potential where the potential energy as a function of distance is given by
\begin{equation}
V\left(r_{ij}\right)=\frac{2}{5}\left(1-\frac{r_{ij}}{r_{i}+r_{j}}\right)^{\frac{5}{2}}
\end{equation}
where $r_{ij}$  is the distance between particle $i$ and $j$, and $r_i$  and $r_j$  are the radii of particles $i$ and $j$ respectively~\cite{johnson1987contact}. We study 50:50 mixtures of particles with a size ratio of 1:1.4 in order to suppress crystallization~\cite{OHern2003}. Two-dimensional systems are initialized with random positions in a square periodic simulation box with equal parts small and large particles. The systems are then instantaneously quenched to zero temperature via FIRE energy minimization~\cite{Bitzek2006}.

After the quench process, the systems are strained using Lees-Edwards boundary conditions~\cite{lees1972computer}. \ems{These boundary conditions are periodic with a shift while crossing the top and bottom boundary proportional to the strain imposed on the system.}
\mlm{To identify the onset of instabilities and avalanches, we first} simulate athermal quasistatic shear (AQS) by taking a small shear step and minimizing the total energy of the system using a FIRE minimization algorithm. Since the system is allowed to relax as long as necessary to find an energy minimum after each shear step, \ems{by definition of quasistatic shear }, this approximates a strain rate that approaches zero in large systems.

Following each strain step, the shear stress of the minimized configuration is measured \ems{using the distances and forces between all pairs of particles:
\begin{equation}
\sigma_{xy}=\sum_{\braket{ij}} \overrightarrow{r}_{ij,x}\overrightarrow{f}_{ij,y}.
\end{equation}
 }
If the instantaneous change in shear stress is larger than a specified threshold \ems{of $N\times10^{-4}$ where $N$ is the system size. }, which signifies an instability, we use a linear bisection algorithm to identify the precise strain at which the instability occurs~\cite{van2014}. Using this procedure, we are able to isolate the system just before and just after an instability corresponding to a particle rearrangement \ems{within a strain window of $\frac{0.01}{N^2}$ }.

Once we have identified a particle rearrangement event, we then wish to simulate the dynamics of that event. In athermal quasistatic shear, the minimum energy states at the end of an avalanche are usually found using fast algorithms, such as conjugate gradient minimization or FIRE, that do not necessarily correspond to realistic dynamics\ems{~\cite{Bitzek2006}}. To maintain physically reasonable dynamics, we instead minimize energy using a computationally expensive steepest descent algorithm with an adaptive timestep. \ems{The timestep of the minimization is determined throughout the algorithm such that no particle moves more than $1\%$ of the average particle radius in a single frame. Furthermore, if the force in the new configuration after a timestep, $\overrightarrow{F}(t+dt)$, is more than orthogonal to the previous force, $\overrightarrow{F}(t)$, ie. $\overrightarrow{F}(t)\cdot \overrightarrow{F}(t+dt)<0$, the step is reversed and the timestep is halved until the new force is in the correct direction or $\overrightarrow{F}(t)\cdot \overrightarrow{F}(t+dt)\geq 0$. This method is equivalent to a noiseless molecular dynamics simulation in the overdamped limit where the velocity, $\overrightarrow{v}$, is given by the force, $\overrightarrow{F}$, with some damping coefficient, $\Gamma$,
\begin{equation}
\overrightarrow{v}=\Gamma\overrightarrow{F}.
\end{equation}
This damping coefficient determines the timescale of the molecular dynamics simulation. We choose this term to be unity to set the natural timescale where the velocity of the simulation is given directly by the force.}

\subsection{Quantifying plasticity using $D^2_{min}$}
\label{secd2min}
Plasticity in disordered systems is well captured by $D^2_{min}$, a measure of the nonaffine motion~\cite{falk1998dynamics} $D^2_{min}$ compares two configurations of a system over a specified radius, in this case five average particle radii:

\begin{equation}
D^2_{min,i}\left(\overrightarrow{X}_1,\overrightarrow{X}_2\right)=\sum_{j:r_{ij}<5\bar{r}}\left(\overrightarrow{r_{ij}}_2- \textbf{S}_i \overrightarrow{r_{ij}}_1\right)^2,
\end{equation}
where $\overrightarrow{X}_1$ and $\overrightarrow{X}_2$ represent the two configurations being compared, $r_{ij}$ is the distance between particles $i$ and $j$, $\bar{r}$ is the average particle radius, $\overrightarrow{r_{ij}}_1$ and $\overrightarrow{r_{ij}}_2$ are the vectors that separate particles $i$ and $j$ in the first and second configuration respectively, and $\textbf{S}_i$ is the best-fit affine transformation that minimizes $D^2_{min,i}$. \ems{The details of this affine transformation can be found in the ESI$^{\dag}$.}

\ems{As is standard~\cite{falk1998dynamics}, to compute $D^2_{min}$ one must choose a lengthscale for the neighborhood over which the affine and non-affine transformations are computed.  Consistent with previous work, we choose $5\bar{r}$. Smaller lengthscales cause the algorithm to fail as the particle may not be in the convex hull of the neighborhood, while larger lengthscales result in $D^2_{min}$ fields that are more homogeneous and fail to capture localized rearrangements.} \ems{Because the neighborhood around particle $i$ is determined by distance, it can change between the two configurations and therefore we use the union of these two neighborhoods to define the set of particles $j$ for each particle $i$.  Since we use a 50:50 binary mixture the average radius is the average of the radii of the two species.}

 Our goal is to measure instantaneous plasticity over time. Therefore, we measure $D^2_{min,i}$ between two configurations separated by a small time window throughout the minimization. The bursts of localized deformation have a duration on the order of one natural time unit, so we choose to measure the plasticity over a time window, $\Delta t$, of $0.2$ natural time units to obtain good resolution. \ems{Furthermore, by choosing a time window larger than our frame rate, we are able to compute a moving average of the $D^2_{min}$ field over time.} We denote the plasticity measured at time $t$ with
\begin{equation}
D^2_{min,i}(t)=D^2_{min,i}\left(\overrightarrow{X}\left(t-\frac{\Delta t}{2}\right),\overrightarrow{X}\left(t+\frac{\Delta t}{2}\right)\right)
\label{Dmin}
\end{equation}
where $\overrightarrow{X}\left(t'\right)$ is the configuration at time $t'$. This measure is a scalar field measured on each particle over space and time.

\section{Results}

\begin{figure*}[h!]
\begin{center}
\includegraphics[width=\textwidth]{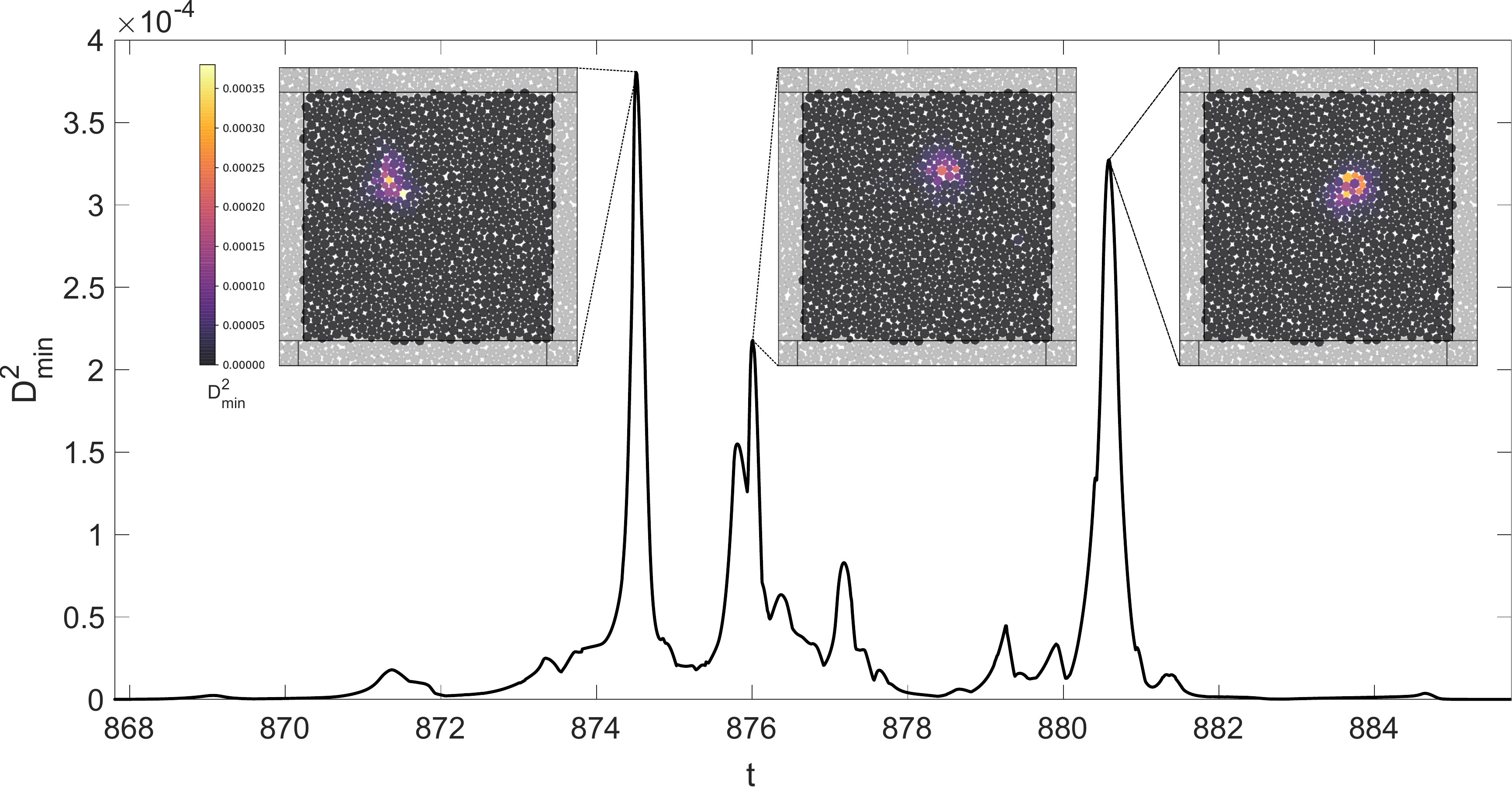}
\end{center}
\caption{Sequential snapshots of the $D^2_{min}$ field are shown in panels (A), (B), and (C). (D) The maximum \ems{$D^2_{min,i}$ over i} as a function of time for an example avalanche \ems{in a system of 1024 particles}. Red symbols indicate the times at which the snapshots were extracted.
}
\label{D2overtime}
\end{figure*}


\subsection{Plastic deformation in avalanches occurs in bursts.}  Examples of the $D^2_{min,i}$ field, defined in Sec~\ref{secd2min}, during one avalanche are shown in Fig. \ref{D2overtime} A, B, and C.

As shown in Fig.~\ref{D2overtime}(D), the maximum value of \ems{$D^2_{min,i}$ over particles $i$} exhibits clear bursts of motion where the maximum value increases by orders of magnitude rapidly and decreases quickly. Furthermore, the $D^2_{min}$ fields shown in panels(A-C) in Fig.~\ref{D2overtime} correspond to the peaks of the three largest bursts of motion. These panels illustrate that the location of these bursts of motion are different for each burst. \ems{Movies which show the spatial extent of the $D^2_{min}$ as a function of time are included in the Supplementary Information.}

For each avalanche, $t=0$ corresponds to the time at which the instability occurs. However, we note that the first burst of localized deformation is often quite delayed. In the example in Fig.~\ref{D2overtime}, the first burst doesn't begin until 868 natural time units after minimization starts. Leading up to that point there is very little motion or activity. This delay occurs because the system begins very near the saddle point instability that triggers rearrangement. Near this saddle point the net force on the system is very small and since the velocity in steepest descent is given by the force, the velocity is also small. It takes time for the system to leave the saddle point behind and approach the time of interest. Similarly, after all the rearrangements have finished, the system relaxes to a minimum and becomes increasingly slow as it approaches. These build-up and relaxation phases take up the bulk of the time during steepest descent minimization, taking on the order of hundreds or thousands of time units, while the system only rearranges for on the order of tens of time units for the system sizes we study \ems{of 1024 particles.}

\subsection{Avalanches can be decomposed into bursts of localized deformation.} It appears that the bursts of localized motion are localized to relatively small groups of particles. To investigate this, we introduce a novel clustering algorithm taking inspiration from persistent homology~\cite{Edelsbrunner2008} and hierarchical density-based clustering methods~\cite{campello2013density}. 

Our goal is to highlight isolated peaks in the nonaffine motion in this system over space and time to quantify whether the motion during an avalanche occurs in localized bursts. The simplest criteria would be a threshold on the nonaffine motion. However, it is clear that applying a bare threshold to a function could easily lose significant peaks and may not well separate the \ems{largest peaks if they are too close together spatially or temporally }. Furthermore, this kind of clustering is very sensitive to the threshold value which must be determined arbitrarily.

\begin{figure}[h!]
\begin{center}
\includegraphics[width=\columnwidth]{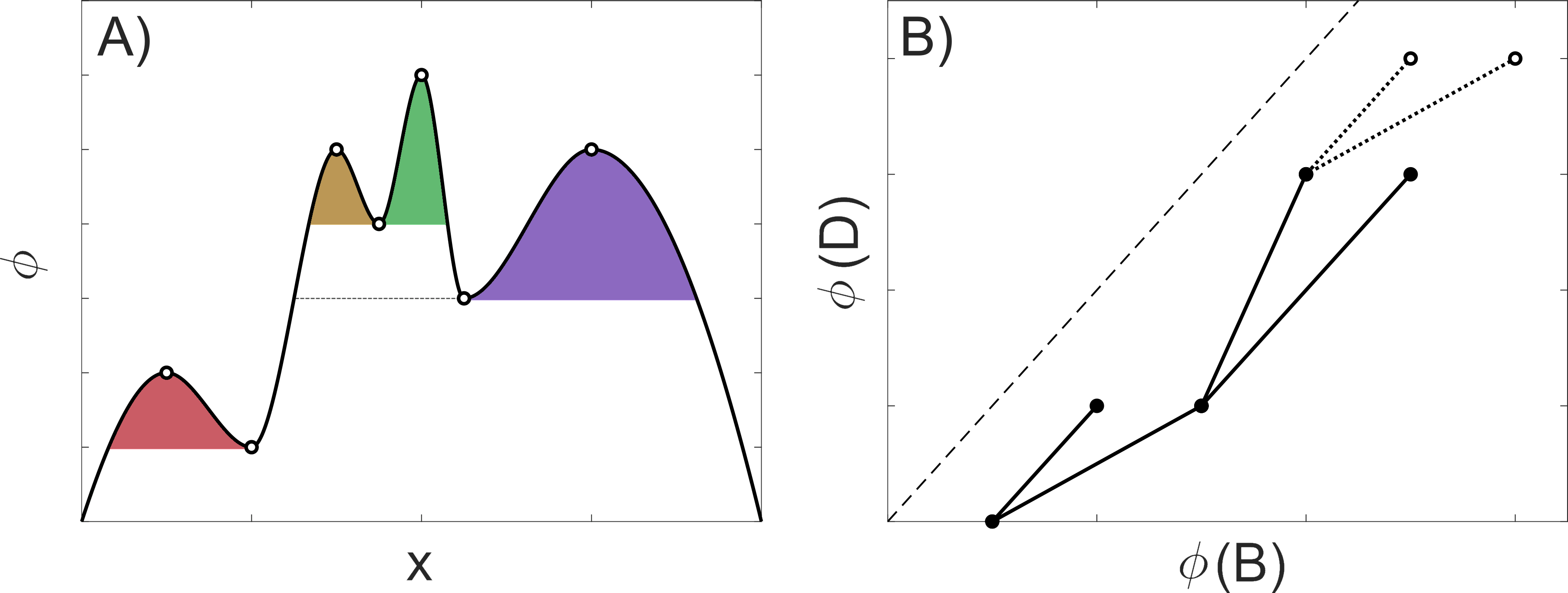}
\includegraphics[width=\columnwidth]{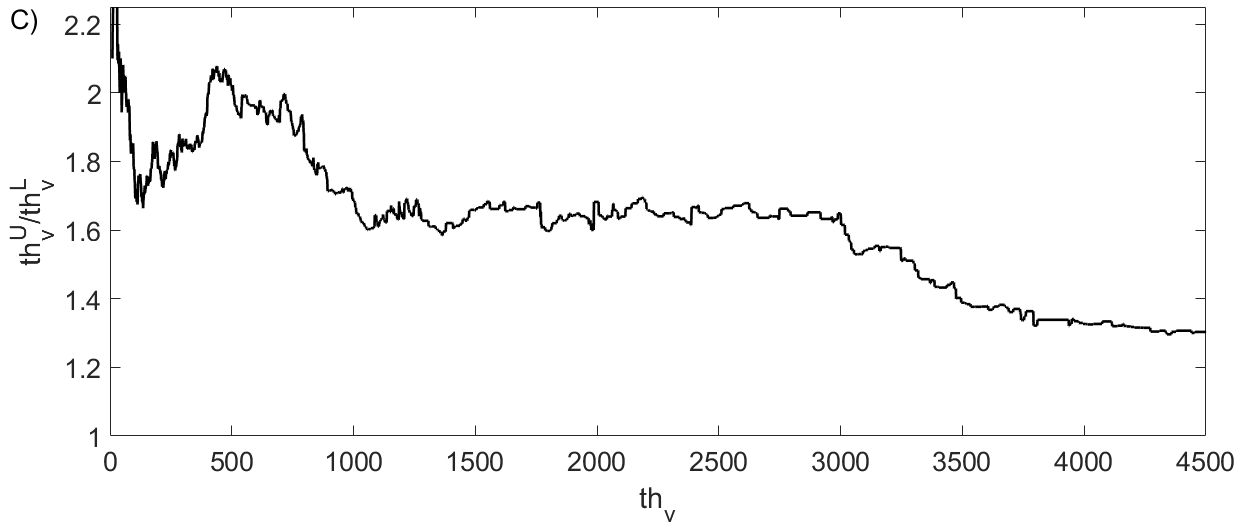}
\end{center}
\caption{
{\bf An extended persistent homology procedure for clustering $D^2_{min}$.}  
(A) A schematic diagram of a function $\phi(x)$ to be clustered using persistent homology. 
(B) The persistence tree for the example field in panel A, where $\phi(B)$ represents the birth height and $\phi(D)$ represents the death height \mlm{of a cluster/peak. Each node in the plot corresponds to a cluster with a volume $v$, and we prune this tree by applying a threshold $th_v$ on cluster volumes.} 
(C) \ems{The ratio $th_v^{U}/th_v^{L}$ describing the 95\% confidence interval around the maximum relative mutual information of cluster fields vs. threshold $th_v$, as described in the main text. Away from zero, this function achieves a broad maximum around $th_v \sim 500$.}
}
\label{Tree}
\end{figure}

By contrast, persistent homology is a sophisticated analysis method for robust characterization of topological features of a set of data or a function over space. It can be used to characterize the height and spatial extent of topological features like local maxima and minima~\cite{Edelsbrunner2008}. This method has been used to quantify the typical heights and sizes of the peaks in a test function and separate them from noise. A schematic diagram of an example test function $\phi(x)$ is shown in Fig.~\ref{Tree}(A), and the corresponding standard persistent homology tree diagram is shown in Fig.~\ref{Tree}(B). 

\ems{In this procedure, a threshold is lowered from the largest value to the lowest value of the function $\phi(x)$. Everything above the threshold value is clustered. When the threshold is lowered below the apex of a peak, a new cluster is formed and the height of the apex is referred to as the birth point, $B$, of the cluster. As the threshold is further lowered, eventually the saddle point which separates 2 or more maxima will be reached and the clusters that were isolated from one another will merge. In typical persistent homology approaches this merging point will mark the "death" of all but one of the previously separated clusters defining the death point, $D$, for these clusters. The typical approach can also result in overlapping clusters, which would be problematic for our task here. Therefore, we take a different approach where a saddle point is marked as the death point of all the merging clusters and the birth point of a new cluster consisting of the merged cluster. This information allows us to construct a tree diagram which consists of the height of birth events $\phi(B)$ and the height of death events $\phi(D)$ with edges connecting the clusters to the clusters they merge to become. The leaves of this tree diagram correspond to the local maxima in the evaluated function.}

\ems{In general, it is expected that some of these leaves are the result of random fluctuations.} To separate signal from noise, one can identify a set of criteria to "prune" the leaves of the tree diagram that correspond to maxima that are simply noise fluctuations, as illustrated by the dotted lines for the two highest branches in Fig.~\ref{Tree}(B). \ems{The remaining leaves identify signals within the clustered function.}


For our problem, we must cluster in space and time simultaneously, which requires that we choose how often to sample in time and set a conversion constant, $c$, between distances in time and distances in space. To ensure good temporal resolution of deformation, we choose a frame rate of $0.01$ natural time units\ems{, which corresponds to a variable number of simulation steps due to the adaptive timestep of the steepest descent method, discussed in Section~\ref{secdyn}}. Next, we define a characteristic length scale for the dynamics of interest, which is roughly the length scale of shear transformation zone also used in the $D^2_{min}$ calculation~\cite{falk1998dynamics}. In 2D, this is about five particle radii, or $l_{char}=5\bar{r}$. Similarly, the characteristic timescale should be roughly the time required for a rearrangement of a shear transformation zone, on the order of a natural time unit. Here we use $t_{char}= 0.1$, \ems{corresponding to 10 frames,} which is half the $\Delta t = 0.2$ chosen for calculating $D^2_{min}$. Then the conversion constant is $c = l_{char}/t_{char} =50$ in units of natural length over natural time. The distance between two particles in space-time is then given by the usual distance with periodic boundary conditions in space, $d(\overrightarrow{x}_i,\overrightarrow{x}_j)$, modified by the temporal distance
\begin{equation}
\tilde{d}(i,j)=\sqrt{d^2(\overrightarrow{x}_i,\overrightarrow{x}_j)+c^2 (t_j-t_i)^2},
\end{equation}
where $\overrightarrow{x}_i$ and $t_i$ are the position and time of particle i. \ems{Following the procedure described above for the test function $\phi(x)$ in Fig.\ref{Tree}A and B, we set a threshold at the maximum value of the $D^2_{min,i}$ field and as the threshold is lowered we cluster all particles with a $D^2_{min,i}$ above this threshold. A particle, $i$, belongs to a cluster if there is any particle in the cluster which is within a distance $l_{char}$ of particle $i$ using the space-time distance measure above. Throughout this process, the birth and death points are noted for each cluster as well as the points which are contained in each cluster at their death point. This information is used to develop a persistence tree diagram like that shown in Fig. \ref{Tree} B.}

We next need to develop a criteria for pruning the tree and separating signal from noise. \mlm{Using our altered approach to avoid overlapping clusters, we found that the standard persistent homology metric for thresholding based on persistence (i.e. the distance of the birth-death coordinate from the diagonal in the tree diagram)} \mlm {did not generate a bimodal distribution of persistence times. Therefore, it was difficult to identify a persistence threshold that separated signal (long-persistence clusters) from noise (short persistence clusters). Instead, we use}
ideas from hierarchical density-based clustering.  Specifically, \mlm{after investigating many possibilities}, we choose to prune the persistent homology tree based on a threshold for the volume of the identified cluster, where the volume is the sum over the number of particles in the cluster \ems{in each frame. A volume of 500, for instance, corresponds to a cluster which may have 5 particles for 100 frames, 100 particles for 5 frames, or some other combination where the number of particles in each frame can vary.} All leaves with volumes below the volume threshold are pruned.  \mlm{The rationale for this approach is that it highlights clusters that are both persistent and involve a sufficiently large region of space-time.} \ems{ To show how this pruning algorithm works in the example of Fig. \ref{Tree}A, the clusters associated with leaves of the tree diagram are highlighted by several colors. If the two central peaks are pruned due to their small size (corresponding to removal of the dashed lines in Fig~\ref{Tree}(B)), the new cluster would correspond to the entire volume above the dashed line in Fig~\ref{Tree}(A). }



\mlm{Next, we must identify the optimal volume threshold. Small clusters that correspond to noise are expected to jump around in space and time as the threshold is changed, while our signal -- localized bursts of deformation -- should be robust with respect to changes in threshold. Therefore, our approach is to calculate the relative mutual information between the clusters identified at different volume thresholds, where mutual information is a measure of the amount of information one variable contains about another. High mutual information means that the clusters do not change much when the threshold is changed, while low mutual information means the clusters change a lot as the threshold is changed. Therefore,} if there is a value of the threshold that generates good separation between signal and noise in this persistent homology representation, we expect to find a plateau in the mutual information, which \ems{would indicate} that the value of the threshold does not strongly impact the clusters found. In other words, all thresholds $th_V$ in the plateau region are sufficient to separate signal from noise.

The mutual information is constructed by measuring the entropy between datasets $I$ and $J$ given by:
\begin{equation}
M(I,J)=\sum_{x\in [I]} \sum_{y\in [J]} p_{x,y} \log_{2}\left( \frac{p_{x,y}}{p_x p_y}\right),
\end{equation}
\ems{where $p_x$, $p_y$, and $p_{xy}$ are the probability of a particle being in set $x$, set $y$, or both simultaneously respectively. In this case the sets $x$ and $y$ are given by the datasets $I$ and $J$, eg. $p_I$ is the probability of a particle being in set $I$ which is computed by the number of particles in set $I$ divided by the total number of particles in the simulation time window, $p_J$ is similarly computed for set $J$, and $p_{IJ}$ is the probability of a particle in both sets computed by the size of the intersect between sets $I$ and $J$ divided by the total number of particles again. }
The relative mutual information between datasets $I$ and $J$ is computed by normalizing by the mutual information by the average \ems{information entropy} between the datasets \ems{which is computed by finding the mutual information of a cluster field with itself}:
\begin{equation}
m(I,J)=\frac{M(I,J)}{\sqrt{M(I,I)*M(J,J)}}.
\end{equation}

\mlm{Our goal is to use the cluster volumes to separate signal from noise, by studying how the mutual information between clusters changes as a function of volume threshold, illustrated in ESI Fig S1$^{\dag}$.
First, imagine an ideal case where the where the clusters that correspond to noise have a lower volume, with a distribution characterized by a mean $th_v^{n}$ and width $dv^{n}$, and clusters that correspond to signal have a larger volume with a distribution characterized by mean $th_v^{s}$ and width $dv^{s}$.  Then a 95\% confidence interval around the maximum value of the relative mutual information will be roughly $dv^{n}$ for smaller values of $th_v$ ($ th_v < th_v^{n} + dv^{n}$) and will change rapidly to a different, presumably larger value $dv^{s}$ for larger values of $th_v^{s}$ ($th_v > th_v^{s} + dv^{s}$).}

\mlm{Therefore, in our data, we search for a threshold at which the 95 \% confidence interval rises rapidly and reaches a plateau, which should correspond to a volume that distinguishes between signal and noise. A subtlety, as discussed the ESI, is that there is also a trend where the} \ems{information entropy, $H(I)\equiv M(I,I)$, in the cluster field, $I$, increases with increasing threshold.} \mlm{To account for this effect, we study the confidence interval of the relative mutual information on a logarithmic scale, which, as discussed in the ESI is equivalent to studying the ratio of the upper and lower contours of the relative mutual information in threshold space $th_v^{U}/th_v^{L}$.}


\mlm{ Fig 2C shows this ratio as a function of threshold volume $th_v$.  There must be a peak at zero as the lower contour is restricted to be positive, and then we see a dip and a rapid increase in the ratio and a maximum around $th_v \sim 500$, followed by a rough plateau.  This indicates that thresholding on cluster volumes indeed separates signal from noise, with an optimal value close to $th_v = 500$, and we use that value for the remainder of this work. This minimum cluster volume corresponds to a localized burst of deformation that contains roughly 10 particles and lasts \mlm{50 frames or }0.5 natural time units. }

\begin{figure}[h!]
\begin{center}
\includegraphics[width=\columnwidth]{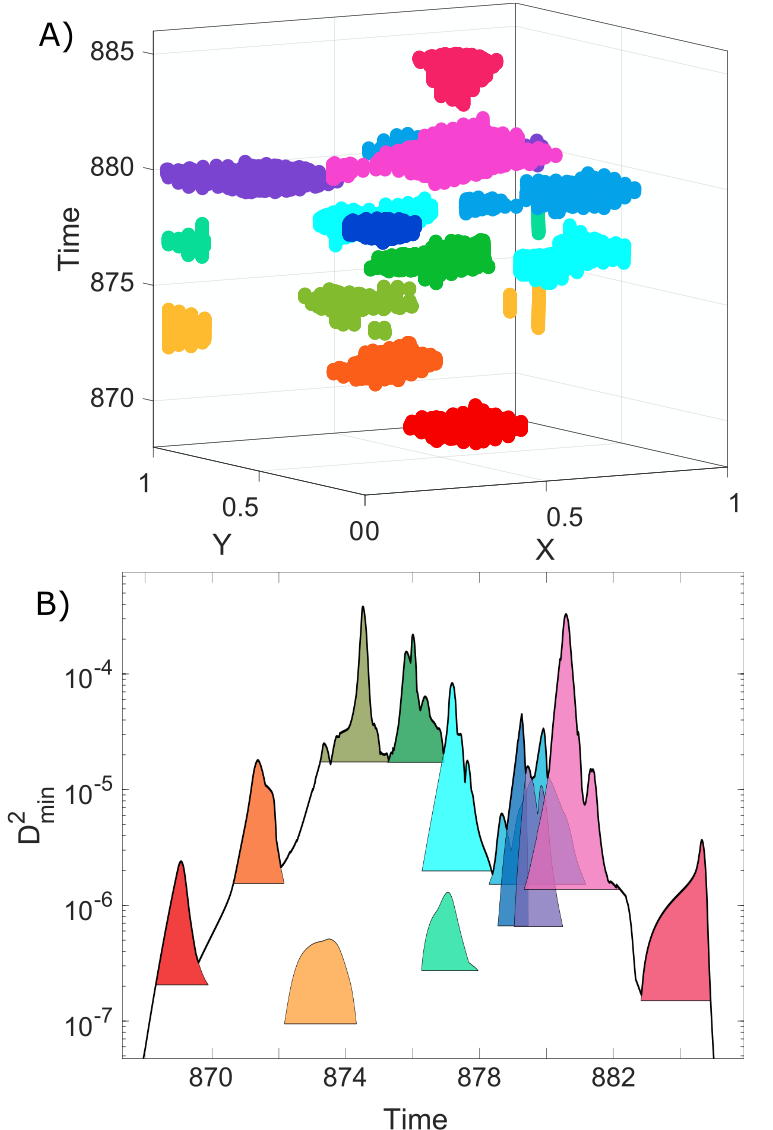}
\end{center}
\caption{(A) A space (X,Y) and time plot illustrating the clusters identified by our persistent clustering algorithm in one example avalanche. Colors indicate unique cluster IDs. (B) \mlm{The black line illustrates the maximum value of $D^2_{min,i}$ over all $i$ at each point in time on a logarithmic scale, for the same avalanche as shown in Fig~\ref{Dmin}. The colored peaks are a plot the maximum $D^2_{min,i}$ within each identified cluster over the period of time the cluster exists. Colors indicate the same cluster ID as in panel A.} 
}
\label{Cluster}
\end{figure}

In Fig.~\ref{Cluster}(A), we show a space-time plot of the clusters of nonaffine motion, as measured by the $D^2_{min}$. Note that this system has periodic boundary conditions in the x- and y- directions, so some of the bursts of localized deformation cross this boundary. \ems{Additionally, because the distance cutoff for clustering is 5 times the average radius, there are some clusters which appear to have gaps in space and time between two or more components. However, it is guaranteed that these gaps are smaller than the characteristic length $l_{char}$.} These clusters meaningfully highlight nonaffine motion in the system during an avalanche. In Fig.~\ref{Cluster}(B), we show the nonaffine motion occurs in peaks over time, where the black curve shows the $D^2_{min,i}(t)$ maximized over \ems{all particles in the system}, indicating that avalanches occur in bursts of motion. The localized clusters on this nonaffine motion are well separated in time and space and represent the local maxima as seen in Fig. \ref{Cluster} B, where the clusters clearly highlight the peaks in motion over time. \mlm{Note that our persistent homology approach allows us to identify significant bursts with relatively small values of $D^2_{min}$, which would have been missed if we instead used an approach with an absolute threshold on $D^2_{min}$ }.

\begin{figure}[h!]
\begin{center}
\includegraphics[width=0.5 \textwidth]{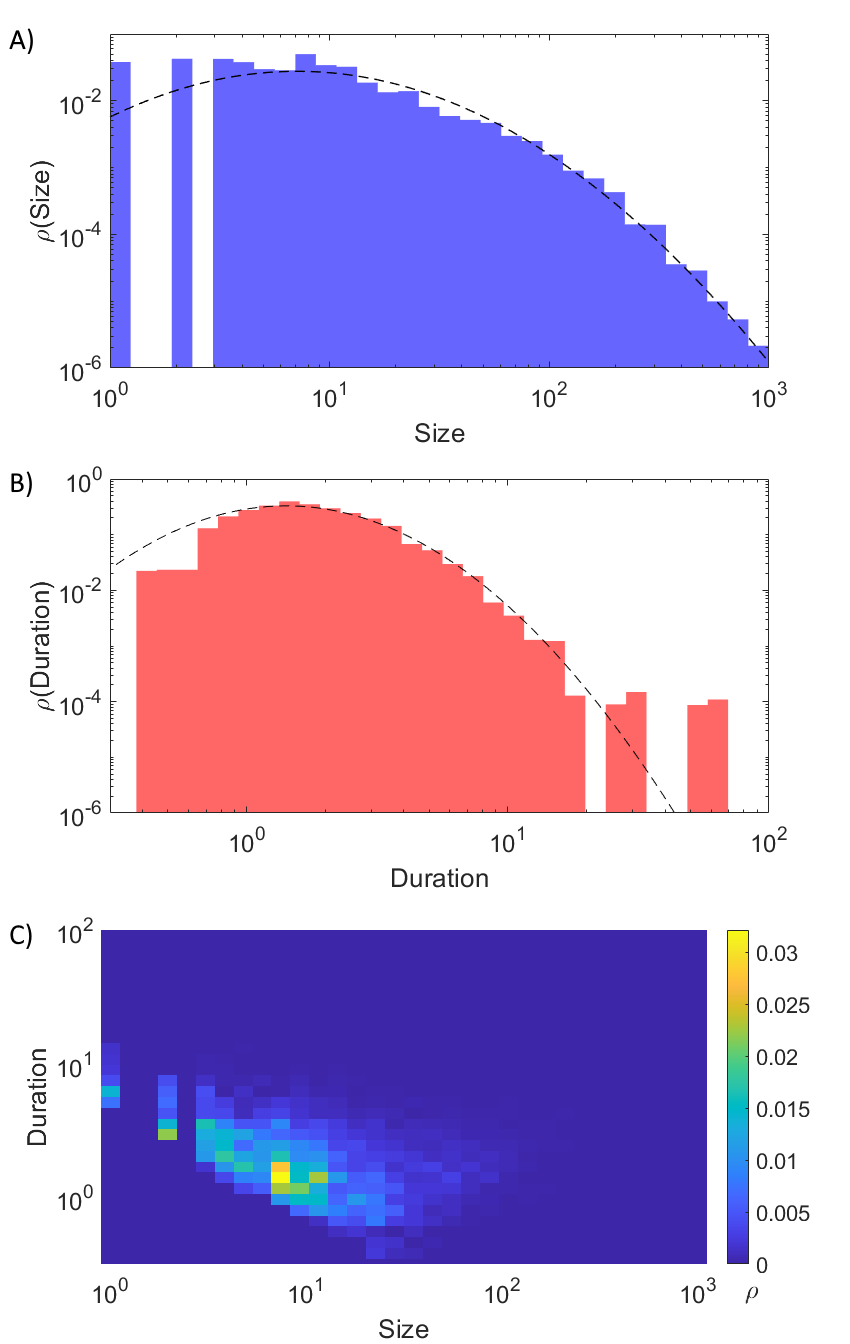}
\end{center}
\caption{ (A) The distribution of the size of the identified bursts of localized deformation in 100 avalanches. The dashed line shows a log normal distribution.
(B) The distribution of the duration of bursts of localized deformation. The dashed line shows a lognormal distribution with the same mean and standard deviation as the duration histogram.
(C) \ems{A histogram where the colorscale indicates the probability density of finding a localized burst of deformation with a given size and duration, identified  over 100 avalanches.}
}
\label{Excite}
\end{figure}

From the beginning of the first burst of localized deformation to the end of the last burst, on average the bursts of localized deformation account for $63\pm 19\%$ of the nonaffine motion while only accounting for $4\pm 2 \%$ of the spacetime volume. These clusters are localized, typically involving less than 100 particles at any given time. The distribution of the spatial extent of the bursts of localized deformation is shown in Fig.~\ref{Excite}(A). This distribution has a heavy tail such that the majority of the bursts are relatively small, where the median of this distribution shows half of the bursts of localized deformation involve fewer than 61 particles. Our data appears to be consistent with a log normal distribution, which is represented by the black dashed line in Fig.~\ref{Excite}(A).

Additionally, we investigate the duration of these bursts of localized deformation. In Fig.~\ref{Excite}(B), we see that the duration of bursts of localized deformation are distributed around unity with a mean value of 2.7 natural time units. Since the duration has a mean that is comparable to the standard deviation but is required to be positive, we hypothesize the distribution of the duration of bursts of localized deformation follows a log-normal distribution, which is shown by the black dashed line and again is consistent with the data.

\mlm{The observation that both distributions are log-normal indicates that individual bursts possess well-defined mean sizes and durations. This is consistent with assumptions of elastoplastic models~\cite{Popovi2018, ozawa2018random} where localized plastic events are coupled via long-range elastic interactions to generate power-law distributions for total avalanche size.} 

Perhaps surprisingly, the duration and the size of each burst of localized deformation do not appear to have a strong correlation, as shown in Fig.~\ref{Excite}(C). In other words, larger bursts do not seem to take longer to complete than smaller bursts of localized deformation.

\subsection{Eigenvalue dynamics in unstable systems are highly complex.}

\ems{In viscously damped inertial dynamics, the equation of motion for a given particle $i$ is given by 
\begin{equation}
\label{inertial}
m_i\overrightarrow{a}_i=\overrightarrow{f}_i-\Gamma \overrightarrow{v}_i,
\end{equation}
where $m_i$ is the mass of particle $i$, $\overrightarrow{a}_i$ is the acceleration of the particle, $\overrightarrow{f}_i$ is the force on the particle, $\Gamma$ is the damping coefficient, and $\overrightarrow{v}_i$ is the velocity of the particle. In overdamped Brownian dynamics, the left-hand side of Eq.~\ref{inertial} is set to zero, under the assumption that the damping term is significantly larger than the inertial term. Therefore, in overdamped dynamics the} velocity is directly proportional to the force . We choose a damping coefficient of unity such that $\overrightarrow{V}=\overrightarrow{F}$\ems{, where each capitalized vector contain $Nd$ entries corresponding to the individual vector components for each particle, where $N$ is the number of particles and $d$ is the number of dimensions}. 

\ems{We compute the forces in this system by taking the derivative of the total energy, $U$, with respect to particle positions, 
\begin{equation}
\overrightarrow{F}_{i\alpha}=\frac{\partial U}{\partial x_{i\alpha}},
\end{equation}
where the subscript $i$ indicates particle $i$ and the subscript $\alpha$ indicates components along the x-axis or the y-axis. We can further compute the change in the force over time to investigate the evolution of structure on short timescales via a time derivative
\begin{equation}
\frac{d\overrightarrow{F}_{i\alpha}}{dt}=\frac{\partial^2 U}{\partial x_{i\alpha} \partial x_{j\beta}} \frac{d x_{i\alpha}}{dt}.
\end{equation}
 The rightmost term is simply the velocity which is equivalent to the force in this simulation and the second order partial derivative of the energy is the Hessian matrix, $H_{ij\alpha\beta}$ . Thus, the change in force is governed by the relation between the force and the Hessian}:

\begin{equation}
\frac{d\overrightarrow{F}}{dt} = - H \overrightarrow{F}.
\end{equation}

When we project the force into the eigenbasis of the Hessian, we find a differential equation for the evolution of the force in the direction of each eigenvector:
\begin{equation}
\frac{d}{dt} F_I(t)=-\lambda_I F_I(t),
\end{equation}
where $F_I=\braket{\overrightarrow{F}|\hat{u}_I}$, and $\lambda_I$ and $\hat{u}_I$ are the $I$th eigenvalue and associated eigenvector respectively. If we assume the eigenbasis of the Hessian doesn't change quickly relative to the timescale of the force, an assumption we will check later, we can integrate this differential equation to find an exponential function over time:
\begin{equation}
F_I(t)=F_I(0)  e^{-\lambda_I t}.
\end{equation}

In mechanically unstable systems where the Hessian possesses at least one negative eigenvalue, the force along these unstable directions grows over time, and the rate of growth depends on the magnitude of the associated eigenvalue.
If the eigenvalues are large the force quickly tracks the eigenvalue, and it is therefore tempting to speculate that the deformation field simply follows the most negative eigenvalue, perhaps with near-instantaneous eigenvalue-switching events, where \mlm{mode associated with} the lowest eigenvalue changes character, due to structural rearrangements and associated contact changes. \ems{Specifically, an eigenmode is said to change character when its direction in configuration space which changes particularly rapidly at these near-instantaneous events. In random matrix theory, these are "narrowly avoided eigenvalue crossings", where the eigenvector associated with the lowest eigenvalue is replaced by a new eigenvector, although eigenvalue repulsion prevents the eigenvalues from actually "crossing".}

\begin{figure}[h!]
\begin{center}
\includegraphics[width=0.5\textwidth]{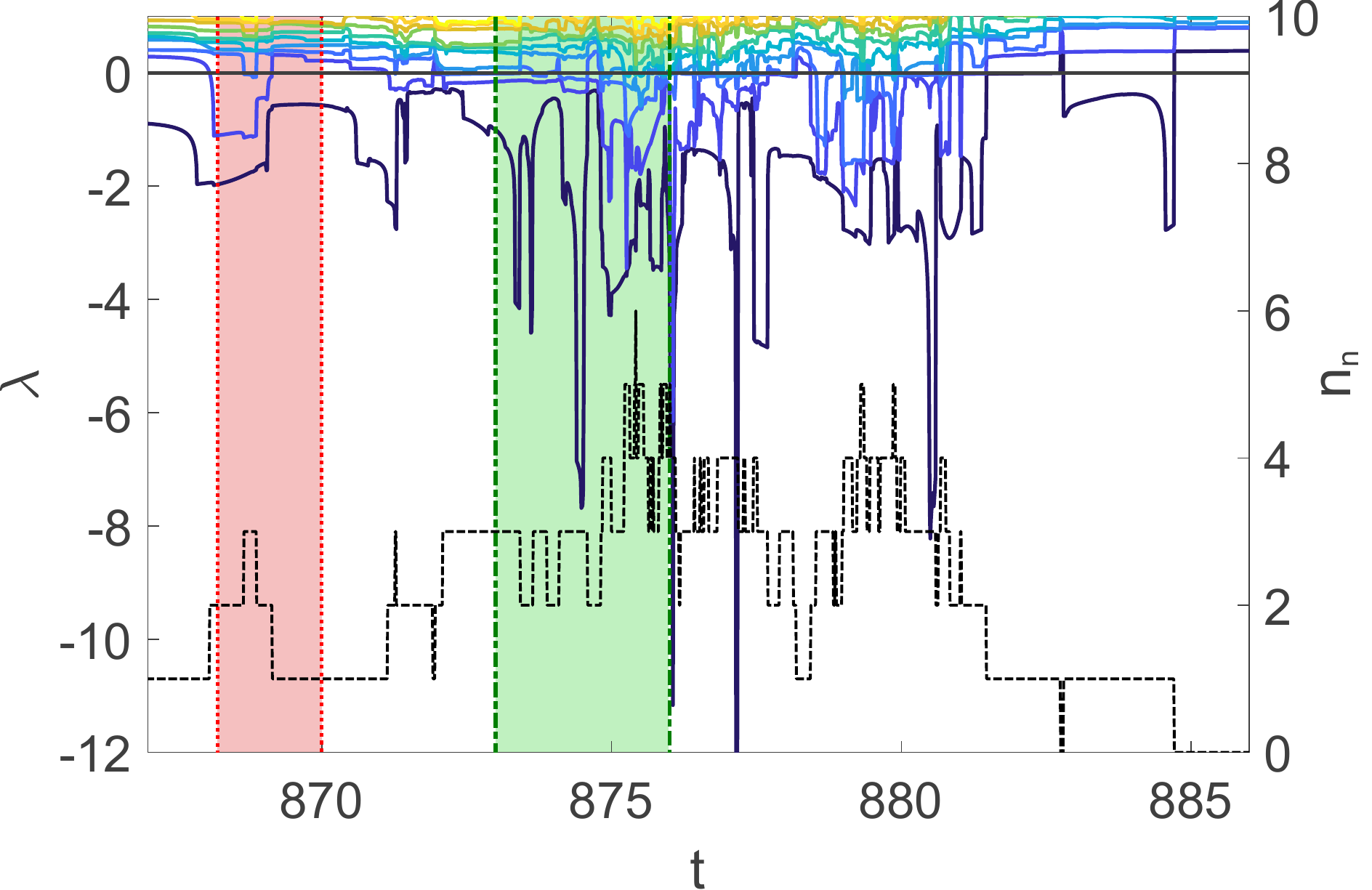}
\end{center}
\caption{ \emph{Left axis:} The lowest 10 eigenvalues of the Hessian as a function of time for a single avalanche simulation shown by solid lines. 
\emph{Right axis:} The count,\ems{$n_n$}, of eigenvalues below zero as a function of time shown by dashed line. 
}
\label{Eigen}
\end{figure}

In our unstable system, we investigate the dynamic behavior of the eigenvalues of the Hessian during deformation in order to probe the curvature of the energy landscape along the minimization path. If the potential energy landscape was simple we would expect a single negative eigenvalue that becomes positive as the system approaches the energetic minimum. In Fig.~\ref{Eigen}, we show the lowest ten eigenvalues over the course of an avalanche. This is the same avalanche example shown in Fig~\ref{Cluster}, and the red and green regions, respectively, correspond to times with bursts of localized deformation with the same color shown in those plots.

Initially, there is only one negative eigenvalue before the main rearrangements occur and, after the entire avalanche is complete, all eigenvalues become positive as the system approaches the minimum, as expected. 

One observation is that, unlike in a simple picture of a single saddle point, many eigenvalues can become negative between the initial configuration, near a saddle point, and the final configuration at a local minimum in the energy landscape. As can be seen in the dashed lined associated with the right axis in Fig. \ref{Eigen}, there can be as many as 5 or 6 negative eigenvalues as the system rearranges. This is indicative of the system passing nearby many saddle points or higher order saddle points during deformation, although our data do not distinguish between these two cases. Another observation is that the eigenvalue dynamics are highly complex, with multiple timescales including very rapid jumps.

To better understand these dynamics and their relationship with the force, we further zoom in on the events associated with the red and green bursts of localized deformation. The red burst is small in magnitude and well-isolated in space and time from other bursts, while the green burst is not.

\begin{figure}[h!]
\begin{center}
\includegraphics[width=\columnwidth]{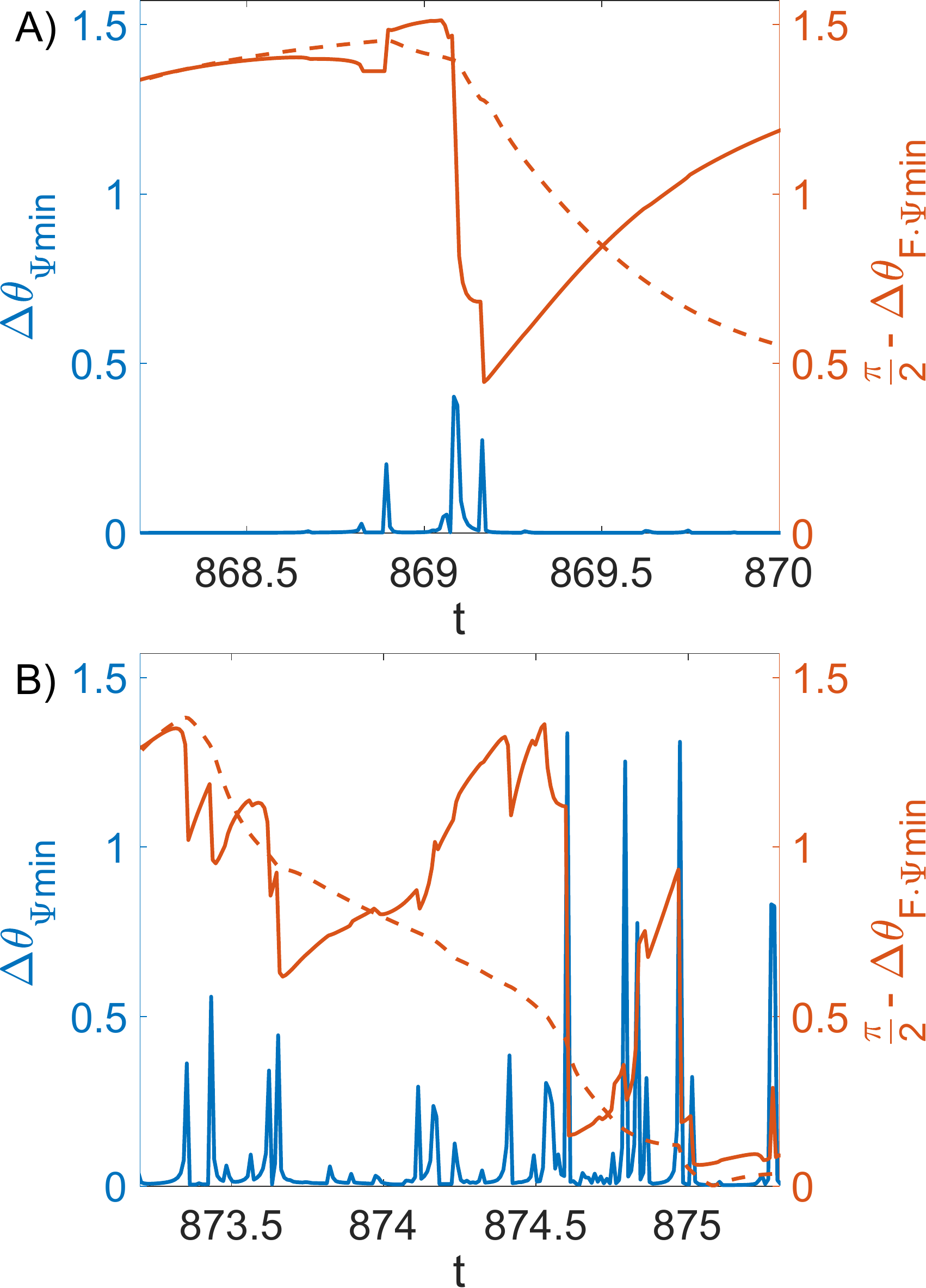}
\end{center}
\caption{ Left axis (blue) The instantaneous angular change in the lowest eigenmode, $\Delta \theta_{\Psi{min}}$, as a function of time.
Right axis (orange) The angle of the force with respect to the lowest eigenmode $\Psi_{min}(t)$, solid; the angle of the force projected into the lowest eigenmode at the beginning of the deformation $\Psi_{min}(t_{ini})$, dashed. 
(A) The eigenvector dynamics during a well-isolated localized burst deformation (the event bounded by the red dotted lines in Fig. \ref{Eigen}).
(B) The eigenvector dynamics during an event with multiple localized deformations simultaneously ( the event bounded in time by the green dashed lines in Fig. \ref{Eigen}).
}
\label{Force}
\end{figure}

In Fig. \ref{Force} (A), the left (blue) axis quantifies the instantaneous change in the angle $\Delta \theta_{\Psi min}$ of the eigenvector with the most negative eigenvalue, which we term the "lowest eigenvector" and denote $\Psi_{min}$ \ems{which directly captures changes to the "character" of the eigenmodes}. Spikes in $\Delta \theta_{\Psi min}$ correspond to "eigenvalue-switching" events, where the lowest eigenvector changes character rapidly, likely due to a change in the contact network.  The right (orange) axis quantifies $ \pi/2 - \theta_{F \cdot \Psi min}$, where $\theta_{F \cdot \Psi min}$ is the angular difference between the force $\bf{F}$ and $\Psi_{min}$.  The dashed (orange) line corresponds to the difference between the instantaneous force and the lowest eigenvector at the start of the burst, $\Psi_{min}(t_{ini})$, \ems{where $t_{ini}$ is determined as the earliest point that the burst has been identified} while the solid orange line corresponds to the difference between the instantaneous force and the current lowest eigenvector $\Psi_{min}(t)$.  When the force is precisely tracking the eigenvector, this quantity is large (close to $\pi/2$), but it approaches zero if the force is orthogonal to the eigenvector.

In this simple isolated burst, we see that are are a handful of rapid, well-separated eigenvalue-switching events. Before the largest such event, the force tracks the lowest eigenvector precisely and the solid orange curve remains close to $\pi/2$. After the largest event, the solid orange curve drops to near zero, indicating that the force is nearly orthogonal to the lowest eigenvalue. Then, over a characterstic timescale governed by the magnitude of the eigenvalue,  the force to begin to track the new lowest eigenvector and the solid orange curve rises again towards $\pi/2$. The dashed line remains low after the eigenvalue switching event, highlighting that the force is no longer tracking the eigenvector that was lowest at the beginning of the burst. Taken together, these data suggest that well-isolated bursts of localized deformation follow our simple sequential picture quite well. 

Fig~\ref{Force}(B) shows the same quantities during the green localized burst.  First, we notice that there are quite a few rapid changes to the lowest eigenmode over time. Moreover, the timescale between such switching events is smaller than the "force-tracking" timescale associated with the magnitude of the eigenvalue, and so the force almost never matches the lowest eigenvalue, as the solid orange curve representing $ \pi/2 - \theta_{F \cdot \Psi min}$ drops and rises several times and and approaches zero by the end of the burst.  Since these more complex events are quite common, it is clear that the lowest eigenmode is not a good predictor of deformation.  This raises the question: are there other indicators based on linear response that would be more stable and therefore more predictive during an avalanche?

\subsection{Soft spots can be identified from a superposition of lowest eigenvalues} As there are rapid changes to the lowest eigenmodes, rather than measuring the overlap with each mode individually, we compute a field similar to the vibrality~\cite{tong2014} over space and time. Vibrality, $\Psi$, quantifies the susceptibility of particle motion to infinitesimal thermal fluctuations in the limit of zero temperature and is proportional to the  Debye-Waller factor.  $\Psi$ is defined as: 
\begin{equation}
\Psi = \sum_{k=1}^{d N-d} \frac{|\Psi_k|^2}{\omega_k^2},
\label{psi}
\end{equation}
where \ems{$N$ and $d$ are the system size  and dimensionality respectively and} the sum runs over the entire set of eigenvectors $\Psi_k$ with frequency $\omega_k$.

To improve performance, we made some alterations to the standard vibrality.  First, as computing the full sum over the entire set of eigenvectors at every time step is computationally intensive, we computed the partial vibrality sums over the $k^*$ modes with the lowest eigenvalues, for varying $k^*$. For stable Hessians and for our system size of interest, we found a partial sum over the lowest eight eigenmodes approximates the true vibrability to within $98\%$. We expect the number of modes needed to  capture the salient features of the structure to increase linearly with system size, though we reserve this for future work.

Second, in the standard vibrality indicator, individual modes are comprised of phonons hybridized with quasi-localized excitations\ems{~\cite{gartner2016nonlinear}.} To remove the phonons, we first compute the non-affine field of each eigenmode by applying the $D^2_{min}$ algorithm on each eigenvector as if it were a displacement field, with the same lengthscale used to quantify deformation, five average particle radii. \ems{In other words, for each eigenmode, we compute $D^2_{min}$ between the current configuration, $X_1$, and a deformed configuration: $X_2=X_1+v_i$, where $v_i$ is the $i$th eigenmode.} Third, because the eigenvalues for unstable Hessians are both positive and negative, and therefore the frequency is not well-defined, we remove the weighting with frequency in Eq.~\ref{psi}. We then calculate a new vibrality-like metric, which we term non-affine vibrality and denote $\widetilde{\Psi}$, as the magnitude of the unweighted sum of the non-affine field associated with the eight lowest eigenmodes of the Hessian. This can be quickly computed at each timepoint $t$ during a steepest descent minimization routine, as it requires only a partial diagonalization of the Hessian, and captures quasi-localized excitations in the eigenmodes of unstable Hessians.

\begin{figure}[h!]
\begin{center}
\includegraphics[width=\columnwidth]{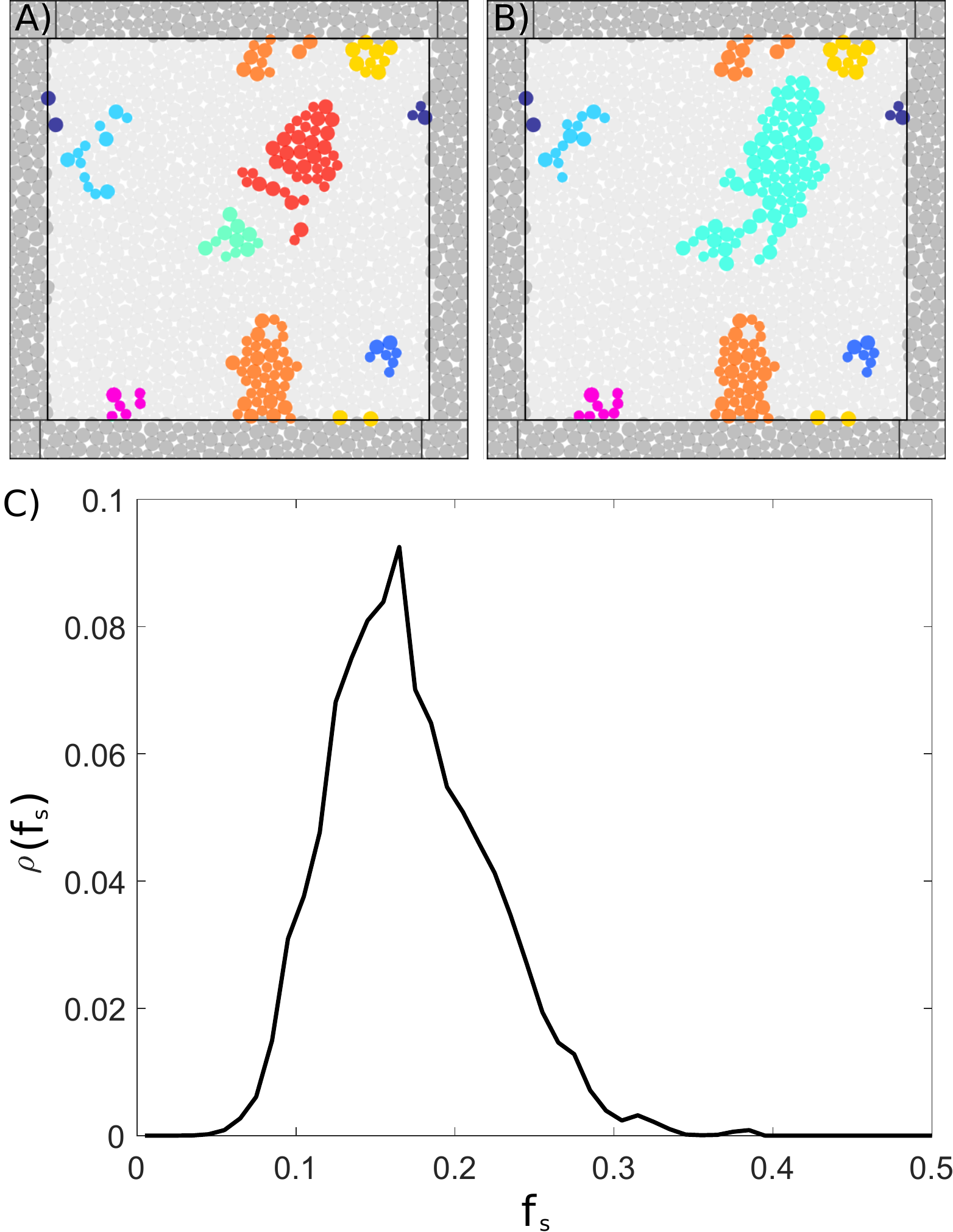}
\end{center}
\caption{(A) The clustered soft spots identified in a single time frame, with periodic boundary conditions in space. Colors index distinct clusters. (B) Soft spots identified one frame later.  (C) Probability distribution $\rho$ of the fraction of the system $f_s$ labeled soft in each frame.
}
\label{SoftSpot0}
\end{figure}

Having identified the non-affine vibrality $\widetilde{\Psi}$ as an efficient structural indicator field for Hessians with negative eigenvalues, we next seek to cluster the field into "soft spots" that can be directly compared with bursts of localized deformations. We found that we were unable to identify soft spots in space and time using the same persistent clustering algorithm used to compute the bursts of localized deformation. This is because while the localized bursts of deformation possess relatively well-defined characteristic length and timescales (see Fig~\ref{Excite}), we observe that soft spots exhibit a very large variation in time: some are very short-lived while others remain for the length of the avalanche. This means there is no choice of volume threshold that separates noise from signal in the persistent homology diagrams.

Although there is no characteristic timescale for soft spots, there is a clear length scale. Therefore, we first use the 
the persistent clustering algorithm to cluster $\widetilde{\Psi}$ as a function of space only at each time slice.  In Fig. \ref{SoftSpot0} A, we show a snapshot of these space-only clusters at a particular time. \ems{Several of these clusters appear to be discontinuous. This is because the clustering lenthscale is $l_{char}$ which is larger than a particle diameter.}
In Fig. \ref{SoftSpot0} B, we show the identified clusters at the next time step. Note that the soft spots near the center of the first frame are joined together in the second frame. To determine how to group these space-only clusters in time, we 
compute the relative mutual information between space-only clusters in adjoining time frames. We expect that spots with large mutual information across time should be grouped together, and those with low information should not.

\ems{In order to separate the information values of objects which overlap and an object that overlaps with the complement of another object,} we use a modified form of the mutual information between two discrete fields computed as 
\begin{equation}
\tilde{M}(I,J)=\sum_{x\in [I,\sim I]} \sum_{y\in [J,\sim J]} p_{x,y} \log_{2}\left( \frac{p_{x,y}}{p_x p_y}\right) \ems{sgn}\left(p_{I,J}-p_I p_J\right),
\end{equation}
where $x$ is the discrete field of the burst of localized deformation, $y$ is the discrete field of soft spots, and $p_x$, $p_y$ and $p_{x,y}$ are the probabilities that an arbitrary point in the discrete fields is $x$ in the discrete field formed from the bursts of localized deformation, or $y$ in the soft spot field, or both, respectively. \ems{Therefore, the term for $x=I$ and $y=\sim J$ investigates the likelihood of particles being in soft spot $I$ but not in soft spot $J$ and all such pairs are studied.} The \ems{Sign function , $sgn$,} in this definition gives positive information if the overlap between $I$ and $J$ is greater than the expected and negative if $I$ is better correlated to the complement of $J$. \ems{ This allows the removal of pairs where the standard measure of the mutual information is high because of the overlap between one set and the complement of the other set.}
The information entropy, $\tilde{H}(I)$, is given by the mutual information of a discrete field with itself, $\tilde{H}(I)=\tilde{M}(I,I)$.  

 \begin{figure}[h!]
\begin{center}
\includegraphics[width=\columnwidth]{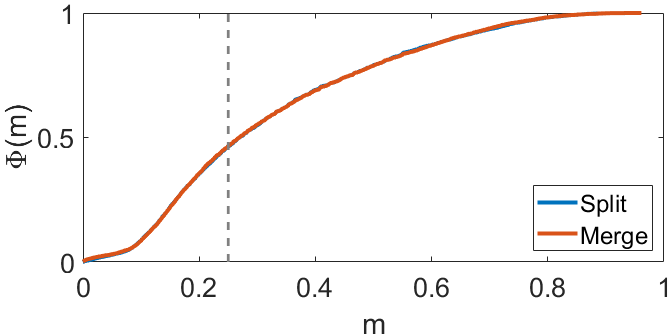}
\end{center}
\caption{The cumulative distribution $\Phi$ of the relative mutual information between soft spots identified in each time frame separated by the time window of $0.1$ natural time units or less averaged over 180 avalanches. These overlaps are separated into cases where the soft spot splits into two soft spots (Split) and where two soft spots merge into one (Merge) which have nearly identical distributions.
}
\label{SoftSpotinfo}
\end{figure}

To determine an information threshold, we analyze the \ems{relative} mutual information probability distribution for simple events where a soft spot in one frame overlaps with two in another frame, shown in Fig. \ref{SoftSpotinfo}. We find that the median value for these types of events is $0.25$, and so we choose that as a threshold. \mlm{As discussed in the ESI$^{\dag}$, this also corresponds to an inflection point between a sharp low-information peak (spots that don't overlap) and a flat high-information distribution (spots that do overlap). We have also checked that varying this threshold slightly does not significantly change our results.}  
In the example shown in~\ref{SoftSpot0}(A) and (B), the upper right overlapping space-only cluster has large enough mutual information to be identified with the space-only cluster in~\ref{SoftSpot0}(B).  \ems{Specifically, the smaller cluster has $17.6\%$ relative mutual information with the blue cluster in \ref{SoftSpot0}(B) while the larger cluster has $49.0\%$ relative mutual information.} This procedure generates from the field $\widetilde{\Psi}$ a discrete set of space-time clusters which we term "soft spots". Fig~\ref{SoftSpot0}(C) is a histogram of the fraction of the system that is labeled as a soft spot in each time step across 150 avalanches, illustrating that our procedure labels about 10-25\% of the system as soft spots, which is consistent with previous methods~\cite{manning2011vibrational}.

\subsection{Bursts of localized deformation occur at dynamically changing soft spots}
To understand how soft spots correlate with bursts of localized deformation, we study the mutual information between these fields~\cite{white2004performance}. Specifically, we use a normalized form of the mutual information called the proficiency, which measures how well each soft spot predicts the spatio-temporal location of each burst of localized deformation:
\begin{equation}
\mathcal{\chi}_{IJ}=\frac{\tilde{M}(I,J)}{\tilde{H}(I)},
\end{equation}
where $\tilde{M}(I,J)$ is the \ems{modified} mutual information between soft spot $J$ and burst of localized deformation $I$ and $\tilde{H}(I)$ is the information measure of the burst of localized deformation $I$. 

\begin{figure}[h!]
\begin{center}
\includegraphics[width=\columnwidth]{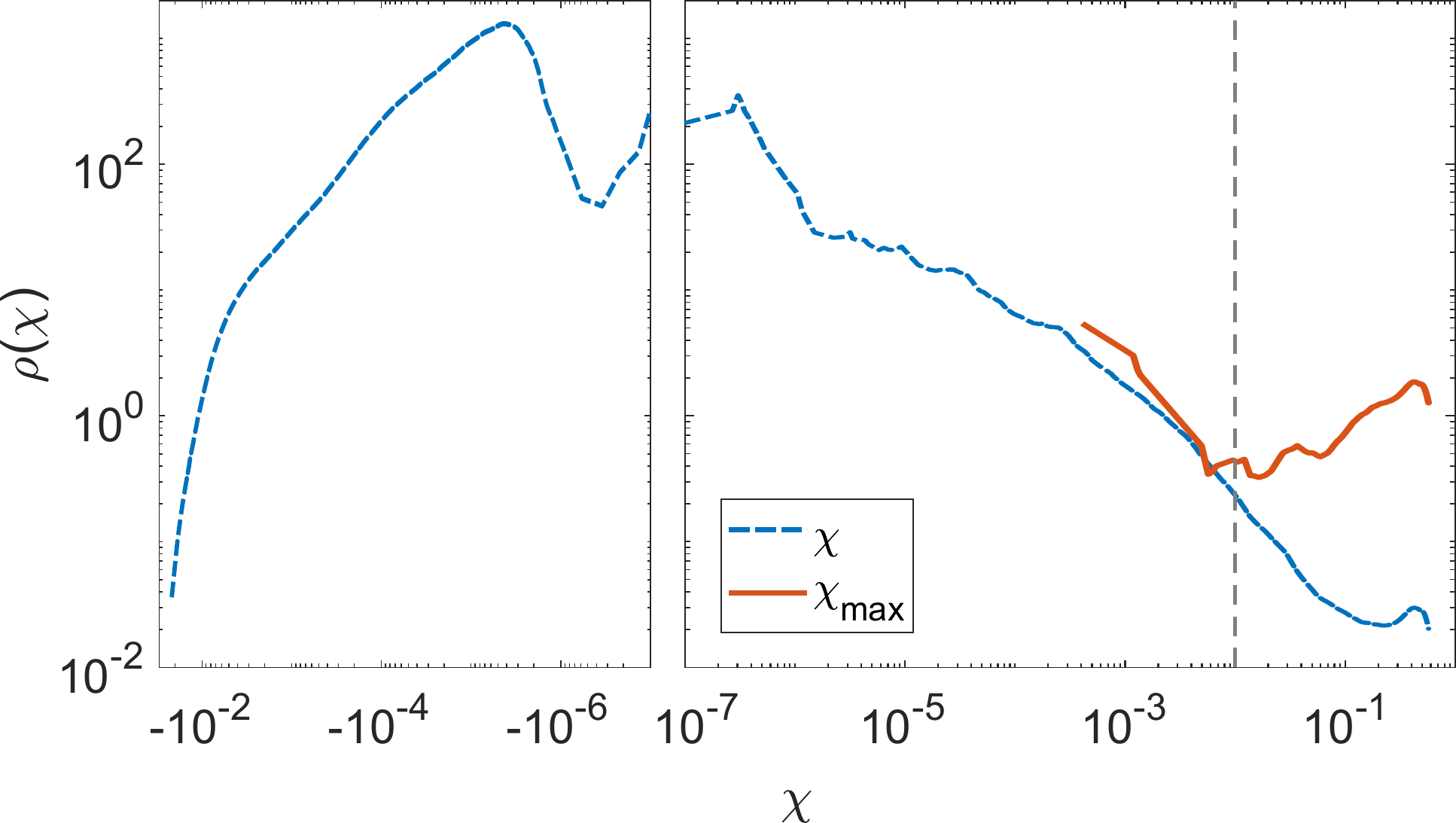}
\end{center}
\caption{Probability distribution function $\rho$ of proficiency $\chi$ between all bursts of localized deformation and all soft spots (blue dashed line), and the maximum proficiency \ems{$\chi_{max}$} for each burst of localized deformation (red solid line), from 180 avalanches.
}
\label{SoftStats}
\end{figure}

The proficiency is near unity when the spatial location of a soft spot overlaps very well with the spatial location of a burst of localized deformation and occurs at the same time. If the proficiency is very near zero, then the soft spot and the burst of localized deformation have little to no overlap in space and time. The probability distribution function for the proficiencies between all soft spots and bursts of localized deformation is shown by the blue dashed line in Fig.~\ref{SoftStats}. The vast majority of proficiencies are very small or negative, indicating that as expected, most soft spots do not overlap with a plastic event at a given instant in time. In contrast, the maximum proficiency for each burst of localized deformation,\ems{$\chi_{max}$} , shown by the solid red line in Fig.~\ref{SoftStats},  exhibits a bi-modal distribution. The peak at high $\chi$ indicates a real overlap between a soft spot and a burst of deformation, while the peak at low $\chi$ is consistent with background noise. Therefore, we define the threshold for overlap at the minimum between these two peaks ($\chi = 0.01$), shown by the vertical dashed line in Fig.~\ref{SoftStats}.

\begin{figure}[h!]
\begin{center}
\includegraphics[width=\columnwidth]{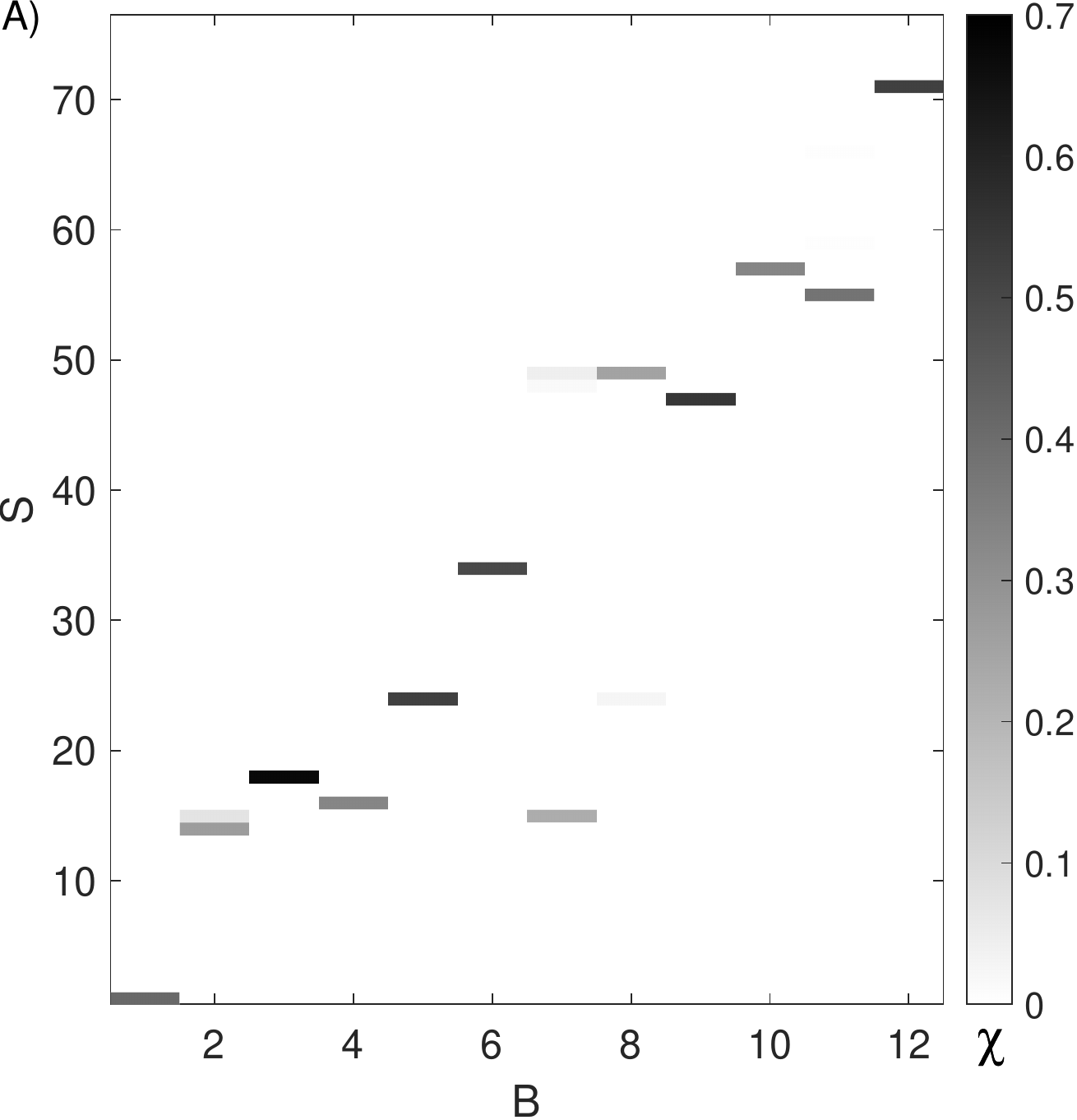}
\includegraphics[width=\columnwidth]{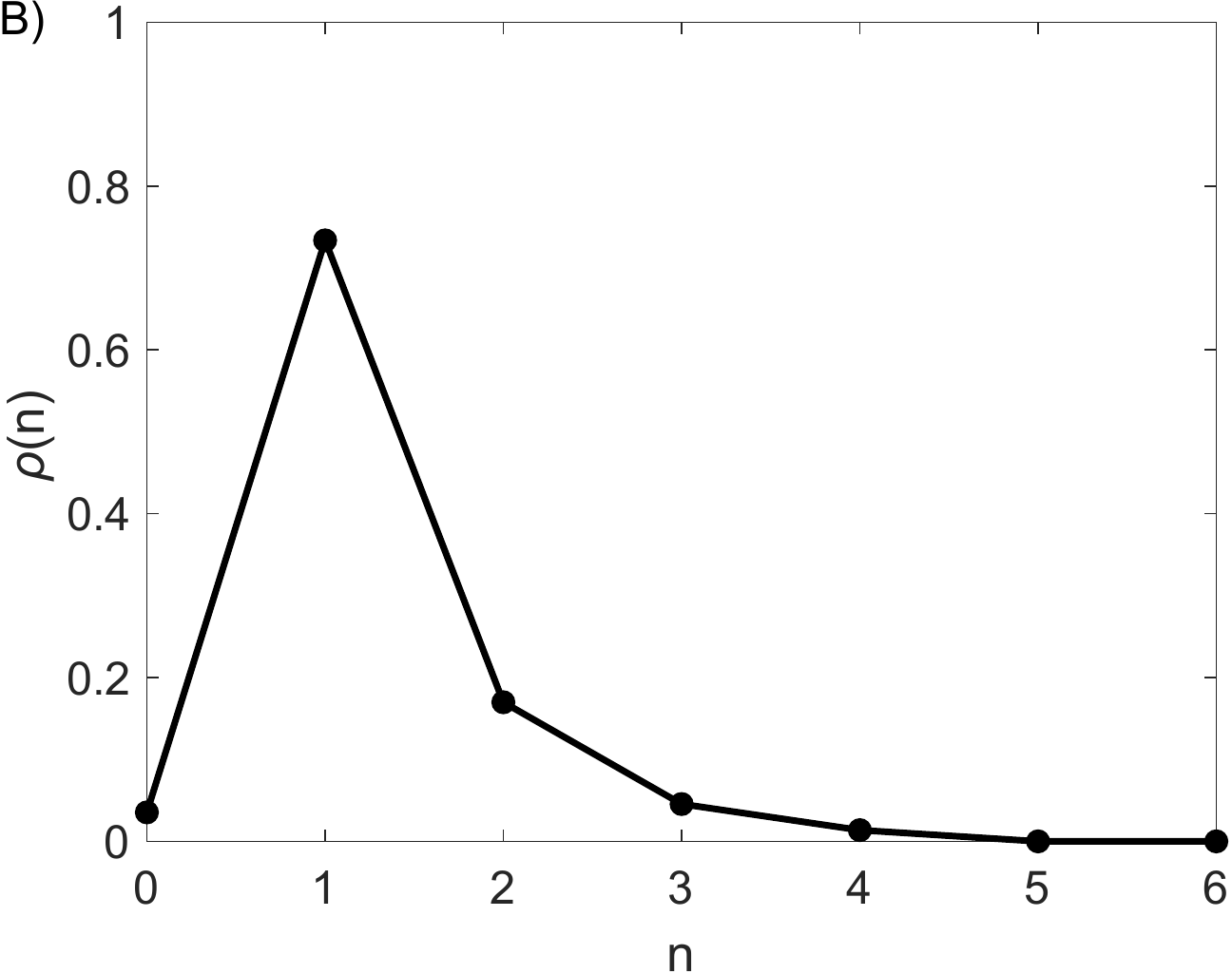}
\end{center}
\caption{(A) {\bf Overlap between 76 identified soft spots and 12 bursts of localized deformation in a single avalanche.} The grayscale colormap indicates the proficiency $\chi$ between each soft spot, indexed by $S$ in order of start time, and each burst of localized deformation indexed by $B$ in order of start time. \textbf{(B) Number of soft spots associated with a single burst of localization.} The probability distribution $\rho$ of bursts of localized deformation that overlap (possess a proficiency $\chi > 0.01$) with $n$ soft spots across 180 avalanches, indicating that most bursts are associated with one soft spots, though a few are associated with 2 or more.}
\label{fig:SoftSpot}
\end{figure}

 Fig.~\ref{fig:SoftSpot}(A) illustrates the relationship between bursts of deformation and soft spots across a single avalanche.  Soft spots are indexed by an integer $S$ in order of their start times, and bursts are similarly indexed by an integer $B$. The colormap indicates the proficiency between each soft spot and a burst of localized information. This example highlights several features that are common to avalanches we studied: i) almost all bursts of deformation are associated with at least one soft spot -- in this example, all bursts of localized deformation have greater than $1\%$ correlation with at least one soft spot, ii) a small number of bursts are associated with more than one soft spot, iii) many soft spots are not involved at all throughout the entire avalanche, iv) some soft spots show up in more than one burst (i.e. those spots rearrange more than once). To get a better idea of the statistics of these features across multiple examples, Fig.~\ref{fig:SoftSpot}(B) shows the fraction of bursts of localized deformation that have an overlap (a proficiency greater than the threshold $0.01$) with $n$ soft spots. Only $3.5\%$ of bursts do not overlap with a predicted soft spot, and the majority overlap with one soft spot. This data is consistent with the hypothesis that bursts of localized deformation occur when a structural defect, or soft spot, reaches its yield stress and deforms.
 
\begin{figure}[h!]
\begin{center}
\includegraphics[width=\columnwidth]{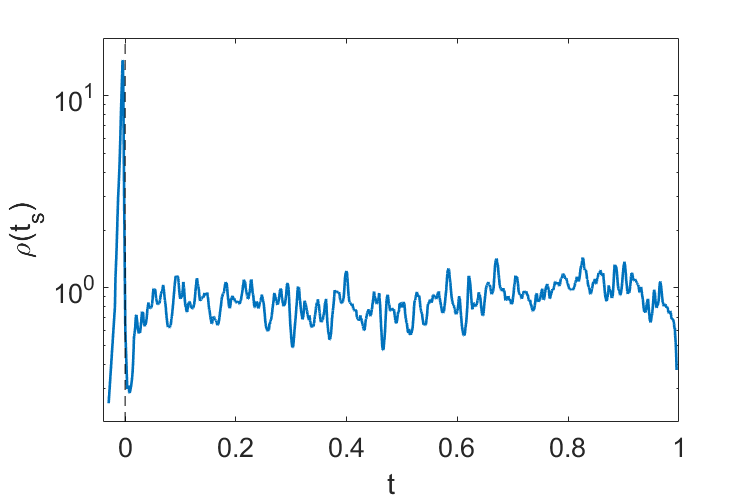}
\end{center}
\caption{The distribution $\rho$ of start times ($t_s$) for soft spots throughout the avalanches where they appear. The start time is normalized such that the instability that triggers the avalanche is occurs at $t=0$ while the avalanche ends and the system finds a new stable state at $t=1$.
}
\label{SoftSpot2}
\end{figure}

 Importantly, many of the soft spots we identify are not present at the start of the avalanche; they appear instead later in the avalanche as a result of the avalanche dynamics. Figure~\ref{SoftSpot2} shows the distribution of soft spot start times normalized by the total duration of the avalanche, averaged over $180$ avalanche trajectories.  While 13\% of the soft spots already exist in the mechanically stable state before the avalanche starts, illustrated by the peak on the left-hand side of~Fig.\ref{SoftSpot2}, the remaining distribution is relatively flat, suggesting that soft spots are equally likely to form at any time during the remainder of the avalanche.
 
\section{Discussion and Conclusions}

In this manuscript, we studied the overdamped dynamics of avalanches in athermal disordered particle packings under applied shear.  Using a set of new persistent-homology-based clustering algorithms, we find that the plastic motion in avalanches generally occurs as sequential bursts of localized deformation. This observation is consistent with elastoplastic theories that explicitly predict avalanches to occur as a sequence of triggered localized rearrangements.

Using the normal modes of the Hessian, we probe the curvatures of the unstable system. One important observation is that there are multiple negative eigenmodes that exchange character extremely rapidly -- on timescales quite a bit shorter than even those localized bursts of deformation. Therefore, the lowest eigenmode is not predictive of the deformation field in such avalanches. Instead,  we develop a vibrality-like structural indicator, $\widetilde{\Psi}$ that is quick to compute, and we cluster this structural indicator field into discrete soft spots that can be directly compared with bursts of localized deformation.

We find that bursts of localized deformation almost always occur at soft spots predicted by our structural indicator. In large avalanches, some of these soft spots are not in the low-energy spectrum at the beginning of the instability, and arise due to the dynamics of the avalanche itself. We find that some bursts are associated with multiple soft spots, and that the same soft spot can sometimes appear in two different bursts, indicating that the same spot rearranges more than once during a single avalanche.

This initial study develops a set of tools that could be broadly useful for analyzing dynamics and structure during avalanches in computer-generated glasses, potentially addressing many long-standing questions in the field.  One important question is whether a rearranging defect triggers the next defect via a simple elastic kernel, as proposed in elasto-plastic theories, via diffusion of structural or effective-temperature like variables, as proposed in Soft Glassy Rheology~\cite{sollich2006soft} and Shear Transformation Zone~\cite{falk1998dynamics} theories, or perhaps a non-trivial combination of such effects~\cite{zhang2020interplay}. By studying how these bursts of localized deformation are coupled in space and time, it should be possible to determine whether they are consistent with elastic propagation at speed set by the elastic moduli. Moreover, it may be possible to determine how the newly formed soft spots are correlated with deformation and determine if that is consistent with theories that postulate diffusion of structural information.

This study has focused on two-dimensional disc packings generated via infinite temperature quench, as visualizing structural defects and their evolution over time is much easier in 2D than 3D.  However, all of the techniques developed here should be fairly straightforward to extend to 3D sphere packings.  In addition, systems prepared via infinite temperature quench are highly ductile and do not exhibit localized shear bands or large stress drops at the yielding transition. Therefore, it will be very interesting to use these new tools to study avalanches in systems that are brittle and where the avalanche is highly localized in space.

Although these methods have been applied to relaxing athermal disordered systems, other unstable systems could also be analyzed using these tools. For instance, studies of the structural evolution of packings in the presence of thermal fluctuations have focused on the inherent, or energy minimized, states or on free-energy minimized configurations.  Similarly, studies of active systems, such as crowds of humans or dense packings of driven colloidal particles, have relied on structural evaluations of mechanically stable reference states. In both cases, instantaneous evaluation of the structure of these mechanically unstable systems was not available. It will be interesting to see if the unstable structural indicators we identify in this work are also predictive of the dynamics there.

\section{Acknowledgements}
We thank Peter Morse for useful discussions. M.L.M and E.M.S. acknowledge support from NSF-DMR-1951921 and Simons Foundation Grant No. 454947. M.L.M acknowledges additional support from Simons Foundation 446222, and E.M.S. from MURI N00014-20-1-2479.



\balance


\bibliography{rsc} 
\bibliographystyle{rsc} 

\newpage

\setcounter{section}{0}

\chapter{\twocolumn[
  \begin{@twocolumnfalse}

\sffamily

\noindent\LARGE{\textbf{Avalanche dynamics in sheared athermal particle packings occurs via localized bursts predicted by unstable linear response}: Electronic Supplementary Information}
  
\vspace{0.6cm}

 \end{@twocolumnfalse} \vspace{0.6cm}

  ]

\section{$D^2_{min}$ Computation}
As described in the main text, the $D^2_{min}$ is a measure of the local non-affine deformation between two configurations, $\overrightarrow{X}_1$ and  $\overrightarrow{X}_2$: 
\begin{equation}
D^2_{min,i}\left(\overrightarrow{X}_1,\overrightarrow{X}_2\right)=\sum_{j\in\partial i}\left(\overrightarrow{X}_{2i}-\overrightarrow{X}_{2j}- \textbf{S}_i \left( \overrightarrow{X}_{2i}-\overrightarrow{X}_{2j} \right)\right)^2,
\end{equation}
where $\partial i$ is the set of particles that are defines as the neighborhood of particle $i$ which we have determined in the main text via the distance from particle $i$ in both configurations and $S_i$ is the best-fit affine transformation around particle $i$ such that any other transformation in place of $S_i$ results in a larger value for the $D^2$ measure. This best-fit affine transformation can be found by calculating
\begin{equation}
X_{i,\alpha\beta}=\sum_{j\in\partial i}\left( \overrightarrow{X}_{2i}-\overrightarrow{X}_{2j} \right)_\alpha \left( \overrightarrow{X}_{1i}-\overrightarrow{X}_{1j} \right)_\beta
\end{equation}
and
\begin{equation}
Y_{i,\alpha\beta}=\sum_{j\in\partial i}\left( \overrightarrow{X}_{1i}-\overrightarrow{X}_{1j} \right)_\alpha \left( \overrightarrow{X}_{1i}-\overrightarrow{X}_{1j} \right)_\beta.
\end{equation}
The best-fit affine transformation is given by $S_i=X_iY_i^{-1}$ \citep{falk1998dynamics}.

\section{Cluster Mutual Information}
In ESI Fig. \ref{AppendixPlot}, we show the relative mutual information between the clusters identified at different size thresholds where the x- and y- axes indicate the volume thresholds of the cluster sets to be compared, and the color indicates the relative mutual information. We highlight using black lines the contours which indicate the size thresholds where this mutual information decreases below $95\%$. Note that the relative mutual information always takes a value between $0$ and $1$.

\ems{As discussed in the main text, we would like to identify regions where the confidence interval associated with the maximal mutual information changes rapidly and reaches a broad maximum. A challenge is that the} \ems{information entropy, $H$, of the set of clusters identified at a particular threshold increases} \ems{with increasing threshold as shown in ESI Fig 1 (B).} \ems{This information entropy is found by computing the mutual information of a field with itself or computing the shannon entropy: 
\begin{equation}
H(I)=-\sum_{x\in [I]} p_{x} \log_{2}\left(p_{x}\right).
\end{equation}
}
\ems{To find an balance between a wide confidence interval for the relative mutual information while still maintaining} \ems{low information entropy -- with low information entropy at smaller threshold values --} {we study the ratio $th_v^{U}/th_v^{L}$  between the value of the threshold associated with the upper curve $th_v^{U}$, highlighted in magenta in Fig S1(B), and the value of the threshold associated with the lower curve $th_v^{L}$, highlighted in green in Fig S1(B). This is equivalent to the width of the confidence interval for the relative mutual information on a log scale. This ratio is shown in the main text in Fig 2(C). }

\begin{figure}[h!]
\begin{center}
\includegraphics[width=\columnwidth]{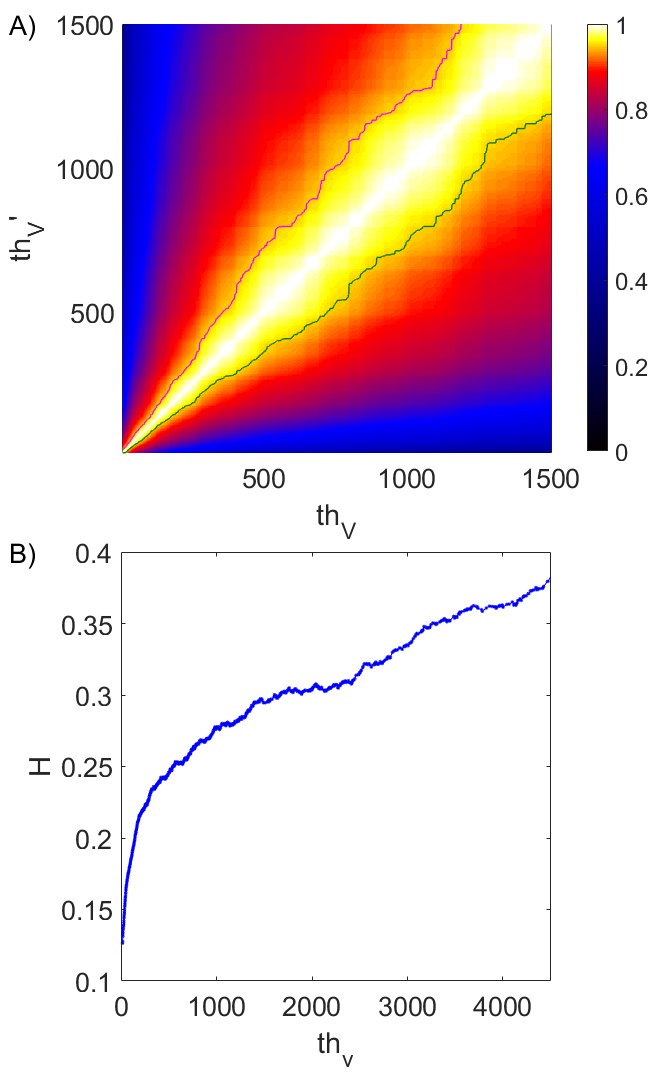}
\end{center}
\caption{ A) The colorscale represents the relative mutual information between the clusters identified at one volume threshold $th_v$, x-axis, compared to those at a different volume threshold $th_v'$, y-axis.  \ems{For each volume threshold $th_v$, we identify two other volume thresholds, $th_v^U$ and $th_v^L$, at which the relative mutual information of the clusters at $th_v$ compared to the clusters $th_v^U$ and $th_v^L$ decreases to 95\%. We order these thresholds such that $th_v^L<th_v<th_v^U$ . The contour generated by this upper threshold, $th_v^U$, as a function of volume is plotted with the pink curve, while the contour generated by the lower threshold, $th_v^L$, is plotted with the green curve. B) The information entropy of the sets of clusters identified at each volume threshold $th_v$.}
}
\label{AppendixPlot}
\end{figure}


\section{Soft Spot Temporal Overlap}

\begin{figure}[h!]
\begin{center}
\includegraphics[width=\columnwidth]{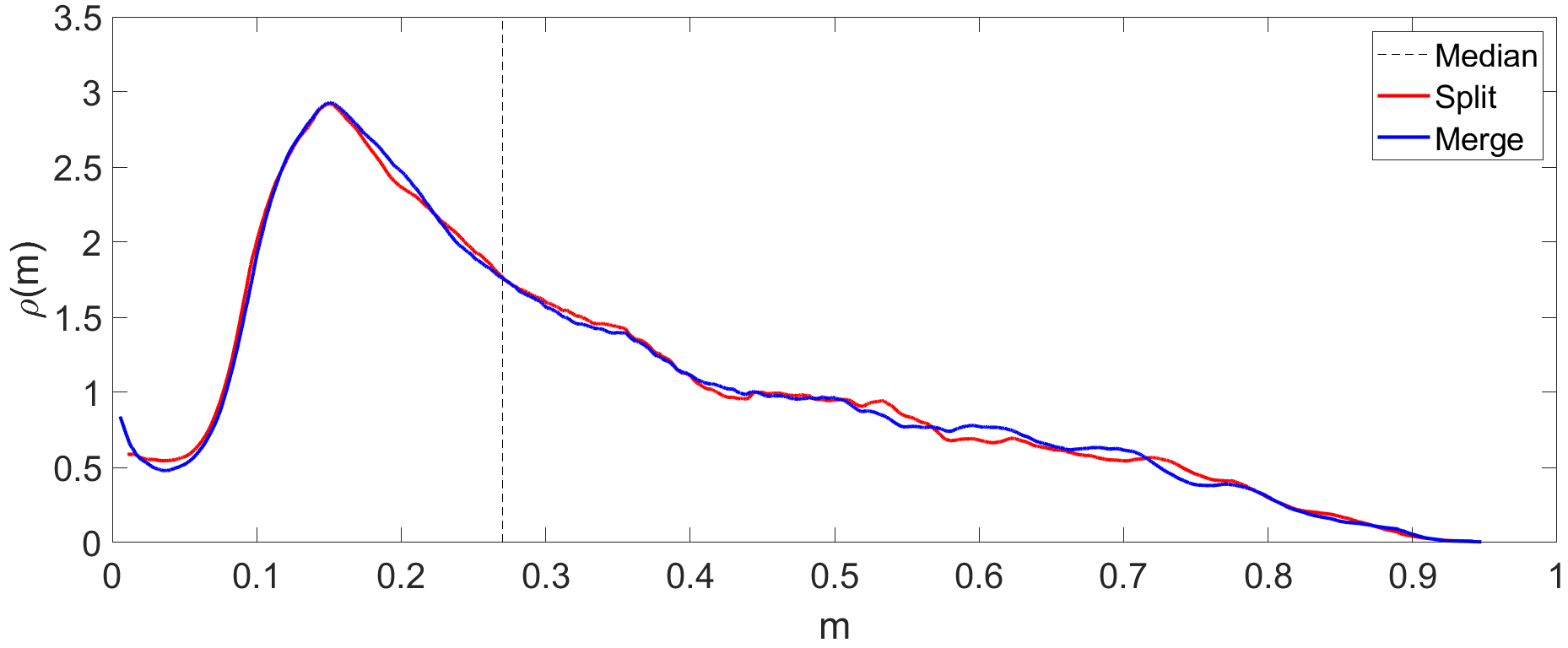}
\end{center}
\caption{The probability distribution $\rho$ of the relative mutual information between soft spots identified in each time frame separated by the time window of $0.1$ natural time units or less averaged over 180 avalanches. These overlaps are separated into cases where the soft spot splits into two soft spots (Split) and where two soft spots merge into one (Merge) which have nearly identical distributions.
}
\label{SoftSpotinfo2}
\end{figure}

By computing the smooth derivative of the cumulative distribution of  the relative mutual information between soft spots, $m$, shown in Fig. 8, we show the probability distribution of $m$ in ESI Fig. \ref{SoftSpotinfo2}. This smooth derivative has been calculated at each point in the cumulative distribution by selecting all points with relative mutual information within $0.03$ of the point in question and computing the best fit slope. 

One possible interpretation of these data is that there is a peak of low-information $m$-values centered around $m\approx 0.15$ which corresponds to spots that don't overlap much and should not be merged together while there is a broader, nearly flat distribution of higher $m$-values that correspond to soft spots that should be identified and merged together. The “cusp” between these two distributions could be estimated as the black dashed line, which happens to be the median and is the value we chose.


\end{document}